\documentclass[titlepage,12pt]{utarticle}
\usepackage{amssymb,amscd}
\usepackage{floatflt,graphicx} 

\usepackage{color}

\numberwithin{equation}{section}

\newcommand{\homoquot}[3]{#3\!\!\left\backslash\frac{#1}{#2}\right.}

\newcommand{\CE}{\sheaf{E}}

\newcommand{\CR}{\sheaf{R}}
\newcommand{\CV}{\sheaf{V}}

\begin{document}
\preprint{
  UTTG--09--01\\
  RUNHETC--2001--19\\
  NSFITP--01--66\\
  {\tt hep-th/0106262}\\
}
\title{
  {\Huge\sl\color{red} Return} of the Torsion D-Branes
}
\author{
  Ilka Brunner%
    \thanks{Work supported in part DOE Grant DE-FG02-96ER40959.\newline\indent
    \email{ibrunner@physics.rutgers.edu}
    }
    \address{
     Department of Physics and Astronomy\\
     Rutgers University\\
     Piscataway, NJ 08855 USA
    },
  Jacques Distler%
    \thanks{Work supported in part by NSF Grant PHY0071512
      and the Robert A.~Welch Foundation.\newline\indent
      \email{distler@golem.ph.utexas.edu}\newline\indent
      \email{rahul@peaches.ph.utexas.edu}
    }
    \address{
      Institute for Theoretical Physics\\
      University of California\\
      Santa Barbara, CA 93106 USA
    }${}^{,c}$
  and Rahul Mahajan${}^\dagger$%
    \address{
      Theory Group, Physics Department\\
      University of Texas at Austin\\
      Austin, TX 78712 USA
    }
}

\date{June 27, 2001}

\Abstract{
We study D-branes on Calabi-Yau manifolds, carrying charges which
are torsion elements of the K-theory. Interesting physics ensues when
we follow these branes into nongeometrical phases of the compactification.
On the level of K-theory, we determine the monodromies of the group of charges
as we circle singular loci in the closed string moduli space. Going
beyond K-theory, we discuss the stability of torsion D-branes as a function of
the K\"ahler moduli. When the fundamental group of the Calabi-Yau is
nonabelian, we find evidence for new threshold bound states of BPS branes.
In a two-parameter example, we compare our results with computations in the
Gepner model. Our study of the torsion D-branes in the compactification of
\cite{FHSC} sheds light on the physics of that model. In particular, we develop
a proposal for the group of allowed D-brane charges in the presence of discrete
RR fluxes.
}

\maketitle


\section{Introduction}\label{sec:intro}
It is by now well established that D-brane charges are classified
by K-theory. More precisely, for branes on a Calabi-Yau,
the even dimensional holomorphic branes
are determined by $K^0$ theory, whereas the charges of the
branes wrapping middle dimensional special Lagrangian cycles
are given by $K^1$. As one moves around in the closed string
moduli space, the group of charges is expected to vary. In particular,
if one traverses non-contractible loops in the bulk moduli space,
the group of charges comes back only up to an automorphism. K-theory
provides a natural framework to study such monodromies, since
those act within $K^0 $ and within $K^1$. Monodromy actions on
the free part of K-theory $K^0/K^0_{tor}$ ( or $K^1/K^1_{tor}$)
have been studied using mirror symmetry. Restricting to the free part
of K-theory, it was sufficient to study the dependence of
$K^0/K^0_{tor}$ on the K\"ahler moduli and $K^1/K^1_{tor}$
on complex structure moduli.

The investigation of the dependence of the torsion part of
K-theory on the bulk moduli was started in \cite{Brunner-Distler}.
There, examples were presented in which the monodromies acted
trivially on the torsion subgroup $K^\bullet_{tor}$. As predicted in
\cite{Brunner-Distler} this is not the case in general, as
will be shown in this paper in examples. It turns out that both
$K^0_{tor}$ and $K^1_{tor}$ undergo monodromy as we vary the
K\"ahler moduli. Since the monodromies act nontrivially on the torsion
subgroup, it follows that they act nontrivially also on the full K-theory.
Similarly, we will find that $K^0$
undergoes nontrivial automorphisms in the complex structure moduli space.

In addition to D-brane charge, K-theory also classifies the fluxes
of the RR-fields. The presence of torsion classes in K-theory
allows one to turn on discrete RR-fluxes. As a result, one might
expect that the full moduli space of a theory has several
branches, corresponding to turning on different fluxes. In
\cite{deBoer:Triples} it was shown that such fluxes can alter the
spectrum of allowed D-brane charges. We will see an example of
such a phenomenon, though our criterion for which fluxes survive
is different from theirs.

In general, one is also interested in the physics of D-branes
beyond K-theory, such as moduli spaces, brane dynamics and
stability. In section \ref{sec:beyond} we start the investigation
of the physics of torsion D-branes, discussing the stability of
brane configuration carrying torsion charge at various points in
moduli space. Since these questions depend on K\"ahler moduli, we
cannot use large volume concepts everywhere, but have to employ
other techniques, such as boundary conformal field theory, to get
information about the stringy regime.

We then move on to discuss three different examples, each of which
is suitable to highlight certain features of torsion D-branes. In
\S\ref{sec:Beauville}, we study one of the very few examples of a
Calabi-Yau with a non-abelian fundamental group. This example
motivates a conjecture for the general form of the monodromy about
(the mirror of) the conifold locus, which generalizes a previous
conjecture which -- we now see -- holds when the Calabi-Yau is
simply-connected.

In \cite{Brunner-Distler}, we found that, when the Calabi-Yau, $X$, is not
simply-connected, there are stable BPS branes carrying 6-brane charge.
These are distinguished from the usual wrapped 6-brane by a discrete
conserved charge -- a torsion element in K-theory. These branes were associated
to one-dimensional irreps of the fundamental group of $X$. In the example
if \S\ref{sec:Beauville}, we also ``discover" the existence of stable
BPS branes corresponding to higher dimensional irreps of the fundamental group.
In contrast to the previous case, these are \emph{not} distinguished by any
conserved (discrete) charge from a collection of wrapped 6-branes. Nonetheless,
we argue that they must be present as threshold bound states in the multiple
6-brane system in order to account for the monodromies that we find.

The second example, discussed in \S\ref{sec:two-param},
is a model whose K\"ahler moduli space has complex dimension 2.
As usual in studying such two-parameter models, the structure of the
K\"ahler moduli space is rather more complicated than in the one-parameter case,
and it is interesting to see that one can produce a consistent set of monodromies
acting on the K-theory. We do have an advantage in this case; the moduli space
contains a Gepner point. So we can compare the results of our topological
calculations with the results from CFT.

The first two examples are two more examples where the monodromies act trivially
on the torsion subgroup of the K-theory.
The third example, discussed in \S\ref{sec:FHSV}
is a free $\BZ_2$ orbifold of $K3\times T^2$.
This manifold played a major role in the context of
heterotic-IIA duality, and D-branes on it were discussed in
\cite{MajSenIII,Tatar:Non-BPS}. It was shown in \cite{FHSC} that IIA theory
on this manifold, with a discrete RR Wilson line turned on,
has a heterotic dual.

This manifold introduces several new features. Unlike previous cases, the
torsion in $K^0(X)$ is \emph{not}  captured by the flat line bundles on $X$.
Second (and related) is the possibility of turning on a flat, but topologically
nontrivial $H$-field, which leads to D-brane charge taking values in the twisted
K-theory (which we also compute). Third, the monodromies \emph{do} act
nontrivially on the torsion subgroups.

The effect of turning on discrete RR flux is nontrivial. First of all, it
changes the global structure of the moduli space (the moduli space is a finite
cover of the moduli space without the RR flux). Second, it restricts the
spectrum of allowed D-branes. In \S\ref{sec:fluxes}, we make a proposal for
the precise form of this restriction. In \S\ref{sec:Singularities}, we
investigate the physics near various singular loci in the moduli space.
In particular, we see that our proposed restriction on D-brane charges in the
presence of discrete RR flux is precisely what is required to make sense of
the physics near the singularity. This ``explains"  from the Type IIA
perspective why it was necessary to turn on the discrete flux.

\section{Review}
In this section, we collect some of the results of \cite{Brunner-Distler}
which will be useful for our present investigations.

Every Calabi-Yau manifold, $X$, whose holonomy group is $SU(3)$, has a
finite fundamental group, and is the quotient of a simply-connected
Calabi-Yau, $Y$ by a finite group, $G$, of freely-acting holomorphic
automorphisms (which preserve the holomorphic 3-form). The most
familiar constructions of Calabi-Yau manifolds -- as hypersurfaces or
complete intersections in toric varieties -- yield simply-connected
Calabi-Yau manifolds, which are candidates for the covering space, $Y$.
The K-theory of such a $Y$ is torsion-free. So, to find a suitable
Calabi-Yau manifold, $X$, with torsion in its K-theory, we look for a
freely-acting group $G$ to mod out by.

The first problem is to compute the K-theory of $X=Y/G$.
For this, we used a pair of spectral sequences, the Cartan-Leray
Spectral Sequence -- which computes the \emph{homology} of $X$ from
the homology of $Y$ -- and
the Atiyah-Hirzebruch Spectral sequence, which computes the K-theory of
$X$ from the cohomology of $X$. Some very useful discussions of the
AHSS have appeared in the recent physics literature
\cite{Brunner-Distler,DiMoWiI,BeGiSu,Bergman-Gimon-Kol}.

For a Calabi-Yau manifold (indeed, for a 6-manifold with  $H^1(X)=0$),
the AHSS converges at the $E_2$ term,
\begin{equation}
    E_2^{p,q}=\coho{p}{X,\pi_q(BU)}
\end{equation}
where $\pi_{2n}(BU)=\BZ$, $\pi_{2n+1}(BU)=0$. And, after a bit of
computation, one finds that the torsion subgroups of the K-theory (our
main interest) fit
into exact sequences
\begin{gather}\label{eq:Ktorext}
    0\to \coho{4}{X}_{tor}\to K^0(X)_{tor}\to \coho{2}{X}_{tor}\to 0 \\
    0\to \coho{5}{X}\to K^1(X)_{tor}\to \coho{3}{X}_{tor}\to 0
\end{gather}
(note that \coho{5}{X} is pure torsion).

The Universal Coefficients Theorem and Poincar\'e duality determine
\begin{gather}
    \coho{5}{X}_{tor}=\homo{1}{X}_{tor}=(\coho{2}{X}_{tor})^*\\
    \coho{4}{X}_{tor}=\homo{2}{X}_{tor}=(\coho{3}{X}_{tor})^*
\end{gather}
and the homology groups can be determined from the CLSS,
a homology spectral sequence with $E^2$ term
\begin{equation}
   E^2_{p,q}= \homo{p}{G,\CH_q(Y)}
\end{equation}
the homology with \emph{twisted coefficients}.

Define the \emph{coinvariant quotient}, $\homo{2}{Y}_G=
\homo{q}{Y}/\CA$, where $\CA$ is the subgroup of $\homo{2}{Y}$
generated by elements of the form $x-g\cdot x$. Assuming that
$\homo{2}{Y}_G$ is torsion-free, we find the needed homology groups to
be given by
\begin{equation}\label{eq:H1}
   \homo{1}{X}=\homo{1}{G}=G/[G,G]
\end{equation}
and the exact sequence
\begin{equation}\label{eq:H2extensionnew}
    0\to \homo{2}{Y}_G\xrightarrow{\pi_{*}} \homo{2}{X}\to
\homo{2}{G}\to 0
\end{equation}
where $\pi_*$ is the push-forward by the projection $\pi:Y\to X$.
With this, the K-theory of the various examples can be calculated.

One notational point. We will frequently be interested in the class
in $K^0(X)$ corresponding to a D-brane wrapped on a (holomorphic)
submanifold $Y\stackrel{i}{\hookrightarrow} X$. This class is
the push-forward in K-theory and should probably be written as $i_! \CO_Y$.
We will abbreviate this as $\CO_Y$. Some readers will note that this
is the same notation used for a certain coherent sheaf on $X$ -- the
structure sheaf of $Y$.  We hope that any confusion that arises
because of this similarity in notation will prove to be beneficial.

As we move away from large-radius, into the interior of the moduli space, the group of D-brane
charges is no longer given by topological K-theory. Still, as a
discrete abelian group, it is locally constant. When we traverse some
incontractible cycle in the moduli space (circle some singular locus),
the group of D-brane charges comes back to itself up to an automorphism.

These monodromies must satisfy the following properties
\begin{itemize}
    \item They descend to the known action on on
    $K^\bullet(X)/K^\bullet(X)_{tor}$, which can be computed, say,
    using Mirror Symmetry.
    \item They preserve  the skew-symmetric intersection pairing
    \begin{equation*}
        (.,.): K^0(X)\times K^0(X) \to \BZ
    \end{equation*}
    given by
    \begin{equation}
      \begin{split}
          (v,w)&=Ind(\overline{\partial}_{v\otimes\overline{w}})\\
               &=\int_X ch(v\otimes\overline{w})Td(X)
      \end{split}
   \end{equation}
   which annihilates $K^0_{tor}(X)$ and is nondegenerate on
   $K^0(X)/K^0_{tor}(X)$. They also preserve the corresponding
   skew-symmetric pairing on $K^1(X)$.
   \item  They preserve the nondegenerate torsion-pairing
      \cite{AtPaSiI, AtPaSiII,AtPaSiIII,Lott}
    \begin{equation*}
       \langle.,.\rangle: K^0(X)_{tor}\times K^1(X)_{tor} \to \BR/\BZ
    \end{equation*}
    \item They commute with the quantum symmetry of the $Y/G$
     orbifold \cite{Vafa89}. Since $G$ acts freely, there are no
     massless states in the twisted sectors. Still, in the full CFT,
     we have an action
     of the quantum symmetry group, $G_Q$. $G_Q$ is the character group of
     $G$ which, in turn,  is
     isomorphic to $G_{ab}=G/[G,G]$. Any character $\chi$ acts by
     phase rotation of the states in the $g$-twisted sector by $\chi(g)$.
     (If $G$ is nonabelian, the twisted sectors are labeled by
     conjugacy classes; $\chi(g)$ only depends on the conjugacy class
     of $g$.) Such characters correspond to the holonomy of connections
     on flat line bundles, so the quantum symmetry group acts on the K-theory
     by
     \begin{equation}\label{eq:QuantumSymmetry}
         v\mapsto v\otimes \CL
    \end{equation}
    for $\CL$ a flat line bundle. The flat line bundles form a group
    under tensor products, isomorphic to $G_Q$.
\end{itemize}

Some of the monodromies we encountered can be described rather generally.
Near large radius, shifting $B$ by an integral class $\xi\in\coho{2}{X}$
is a symmetry of the CFT, which acts on the D-brane charges as
\begin{equation*}
    v\mapsto v\otimes L
\end{equation*}
where $L$ is the line bundle with $c_1(L)=\xi$.

Another locus, at which one has a bona fide conformal field theory,
but nonetheless finds some nontrivial monodromy is given, for example by
Landau-Ginzburg or orbifold loci. Say one finds that, at some locus,
the quantum symmetry group $G_Q$ is enlarged to $\hat G_Q$. One then
finds that this locus is a $\hat G_Q/G_Q$ orbifold locus in the moduli
space. The monodromy about it has finite order,
\begin{equation*}
    M^k=1
\end{equation*}
where $k=|\hat G_Q/G_Q|$.

At the (mirror of the) conifold locus, certain wrapped D-branes become
massless, giving rise to massless hypermultiplets in the 4-d effective
theory. On $Y$, it is the D6-brane, whose K-theory class is the
trivial line bundle, $\CO$ which becomes massless. By the Witten
effect, circling this locus in the moduli space shifts the charges of
all the other D-branes by
\begin{equation}\label{eq:ConifoldMonodromyY}
   v\mapsto v- (v,\CO)\CO
\end{equation}
In \cite{Brunner-Distler}, we saw that, for \emph{abelian} $G$, all of the
flat line bundles (D6-branes carrying torsion charge) become massless,
and \eqref{eq:ConifoldMonodromyY} should be replaced by
\begin{equation}\label{eq:ConifoldMonodromyX}
   v\mapsto v- |G|(v,\CO)\CO
\end{equation}
For nonabelian $G$, the number of flat line bundles is $|G_{ab}|<|G|$.
But, as we shall see in \S\ref{sec:Beauville}, at the conifold locus
there are other branes, corresponding to
flat bundles of higher rank which also become massless, and
\eqref{eq:ConifoldMonodromyX} is the correct formula even when $G$ is
nonabelian.

More generally, say that at some singular locus in the moduli space there is a
collection of D-branes, $W_i$, which become massless. Provided the $W_i$ are
mutually local {\it i.e.}~ provided $(W_i,W_j)=0$, the monodromy about
this locus is
\begin{equation}\label{eq:ConifoldGeneral}
   v\mapsto v- \sum_i(v,W_i)W_i
\end{equation}

This certainly does not exhaust the list of possible classes of monodromies.
One, which will appear in slightly disguised form in \S\ref{sec:FHSV}
is as follows.
Say one has two K-theory classes, $x,y$, with $(x,y)=-2$.
Then let
\begin{equation}\label{eq:KMP}
    v\mapsto v-(v,x-ny)y+(v,y)x
\end{equation}
be the monodromy. This gives rise to extra vector multiplets,
enhancing one of the $U(1)$s to $SU(2)$ (at higher codimension in the
moduli space, one can get higher rank gauge groups), and $n+1$
massless hypermultiplets
in the adjoint (see \cite{Katz-Plesser-Morrison,Klemm-Mayr:Extremal}
for $n\geq1$).
We will see other examples later.

Further insight was gained studying boundary states at the Gepner
point. Building on \cite{BDLR, RecSch}, a set of A-type and B-type
branes for the orbifold X was found. Relative to the states at the
Gepner point of the covering space, these states have extra
labels, corresponding to the torsion charge. By comparing the
intersection form (the Witten index in the open string sector,
$tr_R(-1)^F$) and the action of the quantum symmetry with the
geometrical calculations, the K-theory classes of these branes
could be explicitly identified. By stacking these branes together,
it was possible to construct the torsion branes at the Gepner
point.

\section{Beyond K-theory}\label{sec:beyond}
K-theory, of course, classifies only the conserved charges carried by
D-branes. We would ultimately like to know about the branes
themselves -- which are stable, which are unstable -- and not just
about their charges. In the topologically-twisted theory, a complete
classification of the D-branes themselves is given by the derived category.
Here, one keeps track of
all the massless Ramond fields propagating between a given pair
of D-branes. The ``topological'' D-branes described by the
derived category do not depend on the K\"ahler moduli.
To proceed from topological to physical branes one therefore has
to add in all K\"ahler dependent information, in particular
a notion of stability.

Here, we are interested in physical D-brane systems carrying only
torsion charge. As a first step, we examine those in the context of the
derived category. Moving on to the physical theory, we discuss
the physics at large and small volume separately. As opposed to
the theory of BPS branes, where $\Pi$ \cite{DoFiRoI} and
$\sigma$-stability \cite{OzPaWa} have
been formulated as  stability criteria depending on K\"ahler
moduli, a similar criterion for non-BPS or torsion branes has
not been formulated. Here, we therefore look for tachyon-free
systems, which are stable classically. Beyond that, we try
to find the energetically favorable configurations of a given
charge, arguing that those provide the stable ground states.

\subsection{The derived category}
One important class of torsion branes is constructed as follows.
Given a flat, but nontrivial, line bundle, $\CL$, we can construct
a torsion class in K-theory by subtracting the trivial line bundle
\begin{equation*}
    \alpha=\CL-\CO
\end{equation*}
This cancels the 6-brane charge, and $\alpha$ is a torsion element of
the K-theory. If $\chi=c_1(\CL)$
satisfies $n \chi=0$, then $n'\alpha=0$ for some $n'$ such that
$n$ divides $n'$.

There's an obvious object in the derived category whose K-theory class
is $\alpha$, namely
\begin{equation}\label{eq:obj}
    \{0\to \CO\stackrel{0}{\to}\CL\to 0\}
\end{equation}
a two-term complex, consisting of $\CO$ and $\CL$, with the \emph{zero
map} between them.

In the physical theory, one would like to condense the tachyon
field in the $\overline{D6}D6$ system and reach the ``pure''
torsion brane, which is supported on a tubular neighborhood of the
(torsion) 4-cycle Poincar\'e dual to $\chi$.

In the derived
category approach, one ``understands'' the decay of this $\overline{D6}D6$ system
by first passing to another object which is isomorphic to it in the
derived category (nucleating some brane-antibrane pairs). Then one
deforms that object by turning off certain maps in the complex and
turning on others (turning off and on certain tachyon fields).
One then shows
that the resulting object is isomorphic in the derived category
(quasi-isomorphic in the original category of complexes of coherent
sheaves) to another object which is the end product of tachyon
condensation. The first step (nucleating some brane-antibrane pairs)
is sometimes dispensable.

Of course, whether this decay is
actually favoured energetically is not a question the topological string theory
can answer -- the ``tachyons" are not actually
tachyonic in the topological theory. But, assuming that the decay is
energetically-favoured, the \emph{interpretation} of what happens in
the physical theory depends radically on whether we had to nucleate
some $\overline{D6}D6$ pairs in order to facilitate the decay. If all
we had to do was turn on some tachyon fields, then -- in the physical
theory -- the decay can proceed by classically rolling down the
potential hill. If we had to nucleate some $\overline{D6}D6$ pairs
first (which would  have an energy cost which scaled like the volume of
the Calabi-Yau), then the decay actually proceeds by
barrier-penetration.

Let us apply this procedure to \eqref{eq:obj}. Can we turn on a
nontrivial map $\CO\stackrel{\phi}{\to}\CL$? In the derived
category approach, the answer is \emph{no}. $\CL$ has no
holomorphic sections, so, in the topological theory, there is no
tachyon field, $\phi$, that we can turn on. ($\CL$ does have
meromorphic sections, but those do not correspond to allowed
deformations of the object in the derived category.) Instead, the
decay must proceed by barrier-penetration.

For concreteness, let us work with
the first example of \cite{Brunner-Distler}, which is the quintic in $\BP^4$
modded out by the freely-acting $\BZ_5$,
\begin{equation*}
    (x_1,x_2,x_3,x_4,x_5)\to(x_1,\omega x_2,\omega^2x_3,\omega^3x_4,\omega^4x_5),
    \qquad \omega^5=1
\end{equation*}
The divisor $[x_{k+1}]-[x_k]$ is a representative of the divisor class corresponding
to the nontrivial flat line bundle $\CL$. We start with \eqref{eq:obj}, which
we write as
\begin{equation*}
     \{0\to 0\to\CO\to\CO([x_2]-[x_1])\to 0\}
\end{equation*}
where maps without explicit labels over them denote the zero map. We nucleate a
6-brane-anti-6-brane pair and pass to
the isomorphic object
\begin{equation*}
     \{0\to\CO(-[x_1])\stackrel{c}{\to}\CO\oplus\CO(-[x_1])
     \to\CO([x_2]-[x_1])\to 0\}
\end{equation*}
where $c$ is multiplication by a constant.
Now we \emph{deform} this object by turning \emph{off} the tachyon field $c$,
\begin{equation*}
     \{0\to\CO(-[x_1])\to\CO\oplus\CO(-[x_1])
     \to\CO([x_2]-[x_1])\to 0\}
\end{equation*}
and then turn \emph{on} the tachyon fields $\phi\in \coho{0}{X,\CO([x_1])}$ and
$\phi'\in\coho{0}{X,\CO([x_2])}$,
\begin{equation*}
     \{0\to\CO(-[x_1])\stackrel{\phi}{\to}\CO\oplus\CO(-[x_1])
     \stackrel{\phi'}{\to}\CO([x_2]-[x_1])\to 0\}
\end{equation*}
Finally, we condense these tachyons and pass to the isomorphic object
\begin{equation}\label{eq:finalobj}
     \{0\to0\to\CO_{[x_1]}\to\CO_{[x_2]}\otimes\CL\to 0\}
\end{equation}
The object \eqref{eq:finalobj} has the interpretation of an
anti-D4-brane wrapped on the divisor $[x_1]$ and a D4-brane, with
a flat line bundle on its world-volume, wrapped on the divisor
$[x_2]$. This configuration is supported purely on the divisor
$[x_2]-[x_1]$ and, at least at large radius, it is energetically
favorable for \eqref{eq:obj} to decay to it.

\subsection{Physics at large volume}
While this description of the decay of \eqref{eq:obj} was extremely
pretty (or extremely ugly, depending on your tastes), it is \emph{not}
the dominant decay mode in the physical string theory. In the physical
string theory, at large radius, there \emph{is} a tachyon in the
$\overline{D6}D6$ system, and we can classically roll down the
potential hill. This classical rolling dominates over any decay process,
such as the one described above, which proceeds by barrier-penetration.
This tachyon, however, is \emph{not} related by spectral flow to a
Ramond
ground state, so it does not appear in the topological theory. That is
why the above description of the decay had to proceed by such a
circuitous route -- the relevant tachyon was absent from the derived
category description.

So the derived category is not as helpful as one would like
in studying the decay of
\eqref{eq:obj}.
We  need to work directly in the physical string theory. Let us proceed
physically and estimate the mass of the
tachyon in the $\overline{D6}D6$ system \eqref{eq:obj}. For a \emph{large}
Calabi-Yau, of volume $V$, the mass of the tachyon behaves like
$m^2= -\tfrac{M_s^2}{2}+ \tfrac{c}{V^{1/3}}$, for some positive numerical
constant $c$. This becomes less and less tachyonic as we shrink the
Calabi-Yau. One might wonder whether, at sufficiently small volumes,
this mode ceases to be tachyonic at all. We will turn back to that
question in the next section. For now, we see that the
$\overline{D6}D6$ system will decay due to the presence of the
tachyon. It is energetically favorable to decay to a torsion
$D4$ brane.

We have already seen one such configuration, the $\overline{D4}D4$
system \eqref{eq:finalobj}. Naively, this is a 2-particle state and, in
the topological theory, there is no tachyon that can condense to form
a single-particle bound state. But, since we have already noted that
not all of the potential tachyons in the physical theory have
corresponding fields in the topological theory, we need to look more
carefully.

The divisors $[x_1]$ and $[x_2]$ intersect along a holomorphic curve
in $X$. If the volume of $X$ is sufficiently large, we can treat the
region of intersection of these 4-branes as approximately flat.
Let's say the D4 brane extends in the directions 1256, and the
$\overline{D4}$ in 3456.
The curve along which the divisors  intersect turns into
the 56 plane.  Since the number of ND+DN directions
is $4$, the flat space limit has exactly marginal operators,
parametrizing two branches of the moduli space. In a T-dual setup, this
is like the $\overline{D0}D4$ system, where the Coulomb branch
corresponds to the branes separating and the Higgs branch corresponds
to the $\overline{D0}$ dissolving as an anti-instanton inside the $D4$.
Here the locus of intersection of the branes has real codimension-two
inside the world-volume of one of the branes. The Higgs branch
corresponds to puffing it up to a finite-size vortex.

When the volume of $X$ is not strictly infinite, the exact degeneracy
along these two branches is lifted. The branes are no longer precisely
orthogonal, and so they carry (equal and opposite) charges under the
same RR sector $U(1)$. Thus the degeneracy along the Coulomb branch is
lifted and there is a net attractive force between the branes. The
Higgs branch is harder to analyse, but typically one expects the
vortices of the non-supersymmetric 2+1 dimensional gauge theory to have a
characteristic (nonzero) scale size. Thus, one expects to find the
wave function localized on the Higgs branch -- {\it i.e.}~one expects
that these 4-branes form a bound state.

\subsection{Physics near the conifold point}
When there \emph{is} a tachyon in the open string spectrum between the
brane and antibrane, the \emph{depth} of the tachyon potential represents
the binding energy of the bound state that they form. At large radius, the
binding energy is large because the masses of the 6-branes scale like $V$, while
the mass of the 4-brane scales like $V^{2/3}$ (the $\alpha'$ corrections
are small for large $V$). Whether the end-product of the decay of the
$\overline{D6}D6$ system is a $\overline{D4}D4$ pair, or whether the
latter form a single-particle bound state is a more delicate matter,
involving subleading corrections to the above formul\ae.

At sufficiently small volumes, the energetics that lead to these
conclusions can change completely. Near the conifold point, the volume of the Calabi-Yau becomes very
small, and the D6-brane becomes the lightest BPS particle in the spectrum.
At least in the examples
of \cite{Brunner-Distler}, it has been shown that the torsion D-brane is unstable to
decay into a $\overline{D6}D6$ pair, carrying net torsion charge \cite{Jae:F1}
(see also \cite{GopVaf}).
So there is a curve of marginal stability for the torsion D-brane and this
curve contains the conifold point.

\subsection{Physics near the Gepner point}

Let us now go beyond the naive estimate for the tachyon masses at
small volume and investigate the physics of the D-branes using the
methods of boundary conformal field theory. More specifically, we
are considering free $\BZ_N$ orbifolds of theories with $\CN=2$
worldsheet supersymmetry. We will argue that in this context
tachyon-free brane-anti-brane pairs can be found, indicating that
this is a stable two particle system. Since there is no other
charge present in this setup, its stability shows that there is a
net conserved torsion charge. This is of some importance, since it
is not known how to detect the torsion charge of a boundary state
in BCFT by direct measurement, like for example taking the overlap
with a vertex operator.

At small volume, D-branes are described as boundary conditions
preserving an $\CN=2$ worldsheet supersymmetry. B-type boundary
conditions for BPS D-branes can be regarded as Neumann boundary
conditions for the boson representing the $U(1)$ current of the
supersymmetry algebra. Accordingly, BPS D-branes preserving
different space-time supersymmetry are distinguished by ``Wilson
lines'' for that boson, determining the phase of the central
charge of the BPS state. The charges of the fields propagating in
the open string sector are modulo $2\BZ$ given by the difference
of the Wilson lines characterizing their boundary conditions. In
particular, the open string fields propagating between a brane and
an anti-brane (those have anti-parallel central charges) carry
even integer charge. Such configurations do in general allow for
tachyons, since the vacuum has charge 0 and is not removed from
the open string sector.

To find a stable system, an additional projection is needed in
the open string sector. We can find free $\BZ_N$ orbifold where
there are
$N$ branes carrying the same RR charges. The individual branes
are distinguished by representations $\rho_i$ of the orbifold
group. Open strings in the $\BZ_N$ projected theory transform
in the representation $\rho_i^* \otimes \rho_j$. This projection
is therefore suitable to project out the vacuum from the
brane-antibrane system. All that needs to be done is to attach different
representation labels to the brane and anti-brane.

A concrete example which makes this idea work is the small volume
limit (described by a Gepner model CFT) of the orbifold of the
quintic considered in the previous section. A set of basic brane
constituents is given by the rational $L=0$ boundary states. These
branes are labeled by two integers $M=0,2,4,6,8$ and
$M'=0,2,4,6,8$, which can be understood as irreducible
representations of the  $\BZ_5 \times \BZ_5$ orbifold group. One
factor originates in the GSO projection and the other one
represents the additional geometric scaling orbifold. Accordingly,
the quantum symmetry of the model is $\BZ_5\times \BZ_5$,
generated by $g: M \to M+2, \, M' \to M'$ and $h:M \to M, \, M'\to
M'+2$. A tachyon free brane-antibrane system can be found by
taking a fractional brane $(M,M')$ and its $h$-transformed
anti-brane $\overline{(M,M'+2)}$. the GSO-projection in the open
string channel is then on even integer charges and the only
potential tachyon, the vacuum, is removed from the open string
spectrum by the $\BZ_5$ orbifold action. In this way, we found a
tachyon-free two particle system that is stabilized by the
conservation of torsion charge. Since we started with an arbitrary
brane $(M,M')$ there are altogether $25$ configurations of this
type, which are related by symmetry operations.

All of these are two-particle states. We have seen earlier that at
large volume there is a {\it one}-particle state carrying only
torsion charge, which becomes unstable as we approach the
conifold. An obvious question to ask is what exactly the range of
stability for the single particle state is, in particular, if it
includes the Gepner point.

There is an altogether different way to cancel the RR charge
while keeping net torsion charge \cite{Brunner-Distler}:
One can use a $g$-orbit of
$5$ fractional branes (not including anti-branes) with suitable
$h$ dependence, such as
\begin{equation}
|T> = h|B> + \sum_{n=1}^{4} g^n |B>.
\end{equation}
There are $5$ tachyons involved in this configuration of branes,
meaning that it is unstable. The decay product has to carry the
conserved torsion charge. Condensing four of the tachyons would
lead to a brane-antibrane pair; as discussed previously, this
removes the fifth tachyon. An inequivalent decay process is
induced if all five tachyons are turned on at once. This should
lead to a different decay product and it is suggestive that this
is an $h$-invariant single particle state. Processes involving all
five tachyons (or more generally all links in a quiver diagram)
are rather special, and one usually excludes one type of
fractional brane from the discussion. In the discussion of
\cite{DoFiRoII} removing a link from the quiver enabled a map of
the small volume quiver to the Beilinson quiver describing bundles
on $\BP^n$. Exactly {\it which} link was removed singled out a
particular large volume limit and a particular conifold locus.
Also here, we see that the condensation of only four tachyons
produces a stable brane-antibrane pair, which forms the preferred
configuration at the conifold point.

In the derivation of the low-energy theories of combinations of
$L=0$ branes by adapting orbifold techniques \cite{DiaDou} the
presence of all links prevented a consistent assignment of boson
masses. In \cite{DiaDou} it was proposed that this signals the
breakdown of the low energy field theory description. Our analysis
has shown that such processes are relevant for a full
understanding of torsion branes at small volume.

\section{The Beauville Manifold}\label{sec:Beauville}
In \cite{Brunner-Distler}, we were careful not to assume that the
fundamental group  of the Calabi-Yau was abelian. But there are,
in fact, very few known examples  of Calabi-Yau's whose
fundamental group is a \emph{nonabelian} finite group.

The main example is due to Beauville \cite{Beauville}. Let $Q$ be the group of
unit  quaternions,
\begin{equation}
  Q=\{\pm1,\pm I, \pm J,\pm K\}
\end{equation}
with multiplication law
\begin{equation}
\begin{split}
   I J &= K\qquad \text{(and cyclic)}\\
   I^2 &=J^2=K^2=-1
\end{split}
\end{equation}
Let $Q$ act on $\CV=\BC^8$ via the regular representation. This induces an
action of $Q$ on the complex projective space, $\BP^7=\BP[\CV]$. Let
$Y=\BP^7[2,2,2,2]$
be the intersection of four homogeneous quadrics in $\BP^7$. Beauville showed
that it is possible to choose quadrics such that $Y$ is smooth and $Q$ acts
freely on $Y$. The quotient, $X=Y/Q$ is a Calabi-Yau manifold with fundamental
group $\pi_1(X)=Q$.

Let us recall some facts about the group theory of $Q$. First, there is an exact
sequence,
\begin{equation}
   0\to \BZ_2\to Q \to \BZ_2\times\BZ_2\to 0
\end{equation}
where the commutator subgroup of $Q$ is the $\BZ_2$ subgroup, $\{1,-1\}$
and its
abelianization, $Q/[Q,Q]=\BZ_2\times\BZ_2$.

The irreducible representations of $Q$ are as follows. There are four
1-dimensional irreps: the trivial rep $V_1$ and the representations $V_I,V_J,$
and $V_K$. In $V_I$, $\pm1$ and $\pm I$ are represented by $1$ while $\pm J$ and
$\pm K$ are represented by $-1$ (and similarly for $V_{J,K}$). There is also a
2-dimensional representation, $V_2$ by Pauli matrices.

The representation ring is
\begin{equation}
\begin{split}
   V_2\otimes V_2&= V_1\oplus V_I\oplus V_J \oplus V_K\\
   V_\alpha\otimes V_2 &= V_2\qquad \alpha=1,I,J,K\\
   V_I\otimes V_J &= V_K \qquad \text{(and cyclic)}
\end{split}
\end{equation}
The group homology of $Q$ is
\begin{equation}
   \homo{1}{Q}=Q/[Q,Q]=\BZ_2\oplus\BZ_2,\qquad \homo{2}{Q}=0
\end{equation}

The Hodge numbers of the covering space, $Y$, are $h^{1,1}(Y)=1$,
$h^{2,1}(Y)=65$. In particular, $Q$ must act trivially on
$\homo{2}{Y}$. Plugging into the Cartan-Leray Spectral Sequence,
\begin{equation}
\begin{split}
    \homo{1}{X}=\BZ_2\oplus\BZ_2\\
    \homo{2}{X}_{tor}=0
\end{split}
\end{equation}
$\coho{ev}{X}$ is generated by $1,\xi,\chi_1,\chi_2,\eta$ and $\rho$, with
relations
\begin{equation*}
   \xi^2=2\eta,\qquad \xi\eta=\rho
\end{equation*}
The $\chi_i\in \coho{2}{X}$ are two-torsion, $2\chi_1=2\chi_2=0$. The total
Chern
class of $X$ is
\begin{equation*}
   c(X)=1+8\eta-16\rho
\end{equation*}

A basis for $K^0(X)$ is
\begin{center}
\begin{tabular}{|c|c|c|c|c|c|}
\hline
&$r$&$c_1$&$c_2$&$c_3$\\
\hline\hline
$\CO$&$1$&$0$  & $0$ & $0$\\
$\CO_D=H-\CO$&$0$&
$\xi$  & $0$ & $0$\\
$\alpha_1=\CL_1-\CO$&$0$&$\chi_1$&$0$&$0$\\
$\alpha_2=\CL_2-\CO$&$0$&$\chi_2$&$0$&$0$\\
$\CO_C$&$0$&$0$  & $-\eta$ & $2\rho$\\
$\CO_p$&$0$&
$0$  & $0$ & $2\rho$ \\
\hline
\end{tabular}
\end{center}
Here $H$ is the hyperplane line bundle, $\CL_i$ are nontrivial flat line bundles
on $X$. $C$ is a genus-zero curve in $X$ and, by $\CO_C$, we denote the K-theory
class of a 2-brane wrapped on $C$. $\CO_p$ is the class of a D0-brane at a point
$p\in X$.

In the above basis (omitting the $\alpha_i$, which are torsion elements
of the K-theory) the intersection form,
$(v,w)=Ind(\overline{\partial}_{v\otimes\overline{w}})$ is given by the matrix
\begin{equation}
    \Omega=
    \begin{pmatrix}
    0&-1&-1&-1\\
    1& 0& 1& 0\\
    1&-1& 0& 0\\
    1& 0& 0& 0
    \end{pmatrix}
\end{equation}

The quantum symmetry group, $G_Q$, is isomorphic to the abelianization of $Q$
\begin{equation}
   G_Q=\BZ_2\times\BZ_2
\end{equation}
and acts on the D-branes by tensoring with a flat line bundle,
\begin{equation}
   v\mapsto v\otimes \CL_i
\end{equation}

The K\"ahler moduli space is the 3-punctured sphere. About the
large-radius point, the monodromy is generated by
\begin{equation}
    M_{r}: v\mapsto v\otimes H
\end{equation}
At the conifold point, certain branes become massless. From our
previous experience, we expect that these are the flat line bundles.
There are four such line bundles, $\CO,\CL_1,\CL_2$ and
$\CL_1\otimes \CL_2$. These do, indeed, become massless at the
conifold. But, in addition, there's something else which becomes
massless. As we saw above, $\pi_1(X)$ has a 2-dimensional irrep, out
of which one can build a rank-2 flat bundle on $X$. This rank-2 bundle
(a threshold bound state of a pair of 6-branes, if you wish)
\emph{also} becomes massless at the conifold.

Indeed, we conjecture
that this is a general phenomenon.
Given any Calabi-Yau manifold, $X$, whose holonomy group is $SU(3)$ (and not a
proper subgroup thereof), we can always write it as $X=Y/G$, for $Y$ a
simply-connected Calabi-Yau, and $G$ a finite group.
The monodromy about the conifold locus (principal component of the discriminant
locus) is
always of the form
\begin{equation}
    v\to v- \sum_R (v, W_R) W_R
\end{equation}
where the sum is over all irreps, $R$, of $G$ and $W_R$ is the
flat bundle built using the irrep $R$ of $G$. On the level of
K-theory, it is not hard to see that this can be simplified to
\begin{equation}
    M_c: v\to v- |G| (v,\CO) \CO
\end{equation}
This generalizes an old conjecture of Morrison (see \cite{Morrison:GeomMirror}
and \cite{Kont,Horj}).

In the present case, this is
\begin{equation}\label{eq:BeauvillConifold}
    M_c: v\to v- 8 (v,\CO) \CO
\end{equation}
Finally, the monodromy about the third point,
\begin{equation}
    M_h=(M_{r}M_c)^{-1}
\end{equation}
has, in our example, the property\footnote{This is a particular example of
the monodromy about a ``hybrid point" in the moduli space, where the
``hybrid theory" has the structure of  a Landau-Ginsburg
orbifold
fibered over a $\BP^k$. If the LG orbifold has a $\BZ_n$
quantum symmetry, we find, rather generally,
that the monodromy about the hybrid point satisfies
\begin{equation*}
   (M_h^n-1)^{p}=0
\end{equation*}
for some $p$. In particular, repeating the calculation of the
monodromies for the
covering space, $Y$ (where there is no question that the
conifold monodromy is simply $v\mapsto v-(v,\CO)\CO$), one finds that
\eqref{eq:BeauvilleHybrid} also holds for the monodromy $M_h$ in the
moduli space of $Y$.} that its square is unipotent of
index 4,
\begin{equation}
    (M_h^2-1)^4=0
\end{equation}
A more refined characterization is
\begin{equation}\label{eq:BeauvilleHybrid}
    (M_h+1)^4=0
\end{equation}
{\it i.e.}~that $M_h$ has a single Jordan block with eigenvalue $-1$.
Note that the multiplicity 8 in \eqref{eq:BeauvillConifold} was crucial to
obtaining \eqref{eq:BeauvilleHybrid}.

The existence of the flat rank-2 bundle
as a stable single-particle state was not guaranteed by the BPS
condition (it is degenerate with a pair of D6-branes), nor by K-theory
(it does not carry any K-theory charge by which it might be distinguished
from a pair of D6-branes). Nonetheless, we deduced its existence from the
consistency of the monodromies that we compute. This is another
example of how studying the behaviour of string theory near singularities
can shed light on many subtle issues (in this case, on the existence
of certain threshold bound states).

\pagebreak[4]
\section{A Two-Parameter Example}\label{sec:two-param}
In this section we will study an example of torsion D-branes in which the
K\"ahler moduli space is 2-dimensional.
\begin{floatingtable}{
\begin{tabular}{|c|c|c|}
\hline
& $q_1$ & $q_2$ \\
\hline
$z_1$ & $1$  & $0$ \\
$z_2$ & $1$  & $0$ \\
$z_3$ & $1$  & $0$ \\
$z_4$ & $-3$ & $1$ \\
$z_5$ & $0$  & $1$ \\
$z_6$ & $0$  & $1$ \\
\hline
\end{tabular}}
\caption{Homogeneous coordinates for the resolved model}
\label{tab:T}
\end{floatingtable}

The covering space, $Y$, is the toric resolution of a hypersurface
in weighted projective space, $Y=\BP^4_{1,1,1,3,3}[9]$. Resolving
the orbifold singularities of the weighted projective space yields
a smooth toric variety, $\CT$, which can be realized (in the
language of Gauged Linear $\sigma$-Models) by six chiral
multiplets (the homogeneous coordinates of $\CT$), charged under
$U(1)\times U(1)$, with charges given in Table~\ref{tab:T}. Adding
one more field, $p$, of charge $(0,-3)$ and a gauge-invariant
superpotential, $W= p P(z_i)$, we obtain the GL$\sigma$M for $Y$.

To obtain $X=Y/\BZ_3$, we mod out by a $\BZ_3$ action,
\begin{equation}\label{eq:OrbAction}
   (z_1,z_2,z_3,z_4,z_5,z_6)\mapsto
   (z_2,z_3,z_1,z_4,\ex{2\pi i/3}z_5,\ex{4\pi i/3}z_6)
\end{equation}

The K\"ahler moduli space of $Y$ is 4-dimensional. The exceptional
divisor of $\CT$, $[z_4]$, intersects the Calabi-Yau hypersurface
in three disjoint $\BP^2$s, and there is a K\"ahler modulus
corresponding, roughly, to the size of each of the $\BP^2$s. Only
a 2-dimensional subspace (in which each of the $\BP^2$s has the
same ``size") is represented by toric deformations. This subspace
of the moduli space is parametrized by the complexified
Fayet-Iliopoulos parameters of the $U(1)\times U(1)$ gauge theory
in Table~\ref{tab:T}.

Happily, the orbifold projection \eqref{eq:OrbAction} projects out the nontoric
K\"ahler deformations, and the (2-dimensional) K\"ahler moduli space of $X$
coincides with the subspace of toric K\"ahler deformations of $Y$.

\subsection{Phases of the model}\label{sec:phases}
Let us start out with a brief discussion of the phase structure of
Gauged Linear $\sigma$-Model for this
manifold. Actually, we'll describe the Gauged Linear $\sigma$-model for
the covering space, $Y$, understanding that we will have to mod out by
\eqref{eq:OrbAction} by hand.

As described above,
the linear sigma model has gauge group $U(1) \times U(1)$ and
$7$ chiral multiplets $z_1, \dots z_6, p$. A choice of gauge-invariant
(and $\BZ_3$-invariant) superpotential is given by
\begin{equation}
W= p P(z_i)= p(z_1^9 z_4^3 + z_2^9 z_4^3 + z_3^9 z_4^3 + z_5^3 + z_6^3).
\end{equation}
The possible vacuum configurations have to fulfill the D- and
F-flatness conditions:
\begin{equation}
  \begin{split}
F &= |P|^2 + |p|^2 \sum_i \left| \frac{\partial P}{\partial z_i} \right|^2 \\
D_1 &= |z_1|^2 + |z_2|^2 + |z_3|^2 - 3 |z_4|^2 - r_1 \\
D_2 &= |z_4|^2 + |z_5|^2 + |z_6|^2 - 3 |p|^2 -r_2
  \end{split}
\end{equation}
The model has four phases, depending on the values of the parameters
$r_i$. The limit points of each phase lie at the origin of coordinates for
certain readily-defined coordinate patches on the moduli space. We will
discuss the structure of the moduli space and define the coordinate patches
$U_{ij}$ in \S\ref{sec:monodromies}. In the meantime, we just label the phases
by the corresponding patches:

{\it $U_{34}$ Phase:} $r_1 > 0, r_2>0$. The excluded gauge orbits in this case
are the orbits with $\{z_1=z_2=z_3=0\}$ and $\{z_4=z_5=z_6=0\}$. The F-terms
require the vanishing of $P$ and $p$. As a consequence, the low energy
modes in this limit are a nonlinear $\sigma$-model on  the (smooth) Calabi-Yau
manifold.

{\it $U_{13}$ Phase:} $r_1<0, 3 r_2 + r_1>0$. The orbits
$\{z_4=0\}$ and $\{z_1=z_2=z_3=z_5=z_6=0\}$ have to be excluded.
In a generic D-flat configuration, $z_4$ is not zero. The
Calabi-Yau develops a $\BZ_3$ orbifold singularity at the location
of the blown-down exceptional divisor.

{\it $U_{12}$ Phase:} $r_1<0, 3r_1+r_2<0$. To fulfill D-flatness, the orbits
$\{z_4=0\}$
and $\{p=0\}$ have to be excluded. The F-terms require that
$z_1=z_2=z_3=z_4=z_5=z_6=0$. A gauge transformation by $\ex{i\theta q_1}$
leaves $p$ invariant, while rotating $z_4$. A gauge transformation by
$\ex{i\theta'(q_1+3q_2)}$ leaves $z_4$ invariant, while rotating $p$.
We can use these
two $U(1)$ actions to fix the values of $z_4$ and $p$ completely,
so that the vacuum consists of one point. Around this vacuum, there
are fluctuation of the fields $z_1, z_2, z_3, z_5, z_6$. The VEVs for
$z_4$ and $p$ leave unbroken a $\BZ_9$ subgroup of the $U(1)\times U(1)$,
generated by $\ex{2\pi i(q_1+3q_2)/9}$. In addition, we have to
mod out the theory by the $\BZ_3$ action \eqref{eq:OrbAction}.
Altogether, we arrive at a $\BC^5/\BZ_9 \times \BZ_3$ orbifold model.
Taking into account the superpotential, the
resulting model is a $\BZ_9 \times \BZ_3$ orbifold of a Landau-Ginzburg
model. This Landau-Ginzburg model has an IR description in terms of
the Gepner model $(k=7)^3 (k=1)^2$, on which \eqref{eq:OrbAction}
translates into a permutation of the first three minimal model
factors accompanied by a phase multiplication in the two remaining
factors.

{\it $U_{24}$ Phase:} $r_1>0, r_2<0$. The orbits $\{p=0\}$ and $\{z_1=z_2=z_3=0\}$
have to be removed. This phase corresponds to a hybrid phase:
The fields $z_1, z_2, z_3$ parametrize a $\BP_2$, over which the
fluctuations of the fields $z_4, \dots, z_6$ behave like in a
LG theory. The model has to be modded out by \eqref{eq:OrbAction}.

\subsection{The K-theory}
Since $\homo{2}{\BZ_3}=0$, the CLSS tells us that $\homo{1}{X}=\BZ_3$,
$\homo{2}{X}=\BZ\oplus\BZ$. We can write a basis for $\coho{ev}{X}$,
$1,\xi_1,\xi_2,\chi,\eta_1,\eta_2$ and $\rho$.
The ring structure is
\begin{equation}
  \begin{gathered}
    \xi_1^2=\eta_2,\quad \xi_1\xi_2=\eta_1+3\eta_2,\quad
        \xi_2^2=3(\eta_1+3\eta_2)\\
    \xi_i\eta_j=\delta_{ij}\rho
  \end{gathered}
\end{equation}
where $\chi\in\coho{2}{X}$ is 3-torsion, $3\chi=0$.

A basis for $K^0(X)$ is
\begin{center}
\begin{tabular}{|c|c|c|c|c|c|}
\hline
&$r$&$c_1$&$c_2$&$c_3$\\
\hline\hline
$\CO$&$1$&$0$  & $0$ & $0$\\
$\CO_{D_1}=L_1-\CO$&$0$&
$\xi_1$  & $0$ & $0$\\
$\CO_{D_2}=L_2-\CO$&$0$&
$\xi_2$  & $0$ & $0$\\
$\alpha=\CL-\CO$&$0$&$\chi$&$0$&$0$\\
$\CO_C$&$0$&$0$  & $-\eta_1$ & $2\rho$\\
$\CO_E$&$0$&$0$  & $-\eta_2$ & $0$\\
$\CO_p$&$0$&
$0$  & $0$ & $2\rho$ \\
\hline
\end{tabular}
\end{center}
Here $C$ is a genus-zero curve representing the cohomology class $\eta_1$ and $E$
is an elliptic curve representing the cohomology class $\eta_2$.

In this basis (omitting, as always, the torsion class, $\alpha$), the
intersection form is
\begin{equation}\label{eq:Omega2Param}
    \Omega=
    \begin{pmatrix}
    0&-1&-4&-1& 0&-1\\
    1& 0& 1& 1& 0& 0\\
    4&-1& 0& 0& 1& 0\\
    1&-1& 0& 0& 0& 0\\
    0& 0&-1& 0& 0& 0\\
    1& 0& 0& 0& 0& 0
    \end{pmatrix}
\end{equation}

\subsection{The monodromies}\label{sec:monodromies}
\begin{floatingtable}{
\begin{tabular}{|c|c|c|}
\hline
& $Q_1$ & $Q_2$ \\
\hline
$s_1$ & $3$  & $0$ \\
$s_2$ & $-1$ & $1$ \\
$s_3$ & $0$  & $1$ \\
$s_4$ & $1$  & $0$ \\
\hline
\end{tabular}}
\caption{Homogeneous coordinates for K\"ahler moduli space, $\CM$.}
\label{tab:M}
\end{floatingtable}
Now, before discussing the monodromies, we need to explain a bit about the
structure of the K\"ahler moduli space, $\CM$. $\CM$ is, itself, a toric
variety. Again, we can describe it most succinctly by giving the
GL$\sigma$M data necessary to construct it: a $U(1)\times U(1)$ gauge
theory, with charged fields (homogeneous coordinates for $\CM$) listed
in Table~\ref{tab:M}. $\CM$ is constructed by taking the
Fayet-Iliopoulos parameters $\zeta_{1,2}>0$, imposing the D-flatness
conditions
\begin{equation}\label{eq:MDflat}
    \begin{split}
        3|s_1|^2-|s_2|^2+|s_4|^2&=\zeta_1  \\
        |s_2|^2+|s_3|^2=\zeta_2
    \end{split}
\end{equation}
and modding out by $U(1)\times U(1)$ gauge transformations. Note that the
loci $\{s_1=s_4=0\}$ and $\{s_2=s_3=0\}$ are excluded, as one cannot satisfy
\eqref{eq:MDflat} there. Also note that the locus $\{s_4=0\}$ is a $\BZ_3$
orbifold locus in $\CM$, as $s_1\neq0$ leaves an unbroken $\BZ_3$ subgroup of
the first $U(1)$.

$\CM$ can be covered by coordinate
patches $U_{ij}$ in which $s_i$ and $s_j$ are nonvanishing. Each of
these coordinate patches corresponds to a ``phase'' of the GL$\sigma$M
analysis of $X$, which we reviewed in \S\ref{sec:phases}.

The boundaries of the moduli space are the four divisors $[s_i]$ as well as the
``discriminant locus", which has two components, the conifold locus,
$\Delta_0=[s_1s_2^3-\tfrac{1}{27}(s_2s_4-\tfrac{1}{27}s_3)^3]$,
and another locus, $\Delta_1=[s_1-\tfrac{1}{27}s_4^3]$. The
intersections of these divisors are depicted in
Figure~\ref{fig:modspace}. The figure is a bit deceptive. $\Delta_0$ intersects
almost every divisor represented by a horizontal line in the figure
in three points. The exceptions are $[s_1]$, which it meets tangentially, and
the orbifold locus, $[s_4]$. We placed an extra white dot on $\Delta_1$ to
remind you that it intersects twice more (off this real slice) with $\Delta_0$.

The divisors $[s_1]$ and $[s_2]$ correspond, respectively, to the
$r_1\to\infty$ and $r_2\to\infty$ limits on the Calabi-Yau $X$.
The ``large radius limit'' is located at the intersection of these
two divisors. The monodromies about these divisors are
\begin{equation}
  \begin{split}
    M_{r_1}&: v\mapsto v\otimes L_1\\
    M_{r_2}&: v\mapsto v\otimes L_2
  \end{split}
\end{equation}
In the basis of \eqref{eq:Omega2Param}, these are represented by the
matrices
\begin{equation}
    M_{r_1}=
    \begin{pmatrix}
        1 & 0 & 0 & 0 & 0 & 0  \\
        1 & 1 & 0 & 0 & 0 & 0  \\
        0 & 0 & 1 & 0 & 0 & 0  \\
        0 & 0 & 1 & 1 & 0 & 0  \\
        0 & 1 & 3 & 0 & 1 & 0  \\
        0 & 0 & 1 & 1 & 0 & 1
    \end{pmatrix},\qquad
    M_{r_2}=
    \begin{pmatrix}
        1 & 0 & 0 & 0 & 0 & 0  \\
        0 & 1 & 0 & 0 & 0 & 0  \\
        1 & 0 & 1 & 0 & 0 & 0  \\
        0 & 1 & 3 & 1 & 0 & 0  \\
        0 & 3 & 9 & 0 & 1 & 0  \\
        0 & 1 & 6 & 0 & 1 & 1
    \end{pmatrix}
\end{equation}

In specifying the monodromies about more ``distant" divisors in the moduli
space, we need to specify what path we take in circling them. It will be useful
to choose a (real) 2-surface homotopic to one of our coordinate divisors, passing
through our chosen basepoint near large radius, and specify that the path is
constrained to lie \emph{on} that 2-surface.
\begin{floatingfigure}{3.5in}
   \mbox{\includegraphics[width=3.25in]{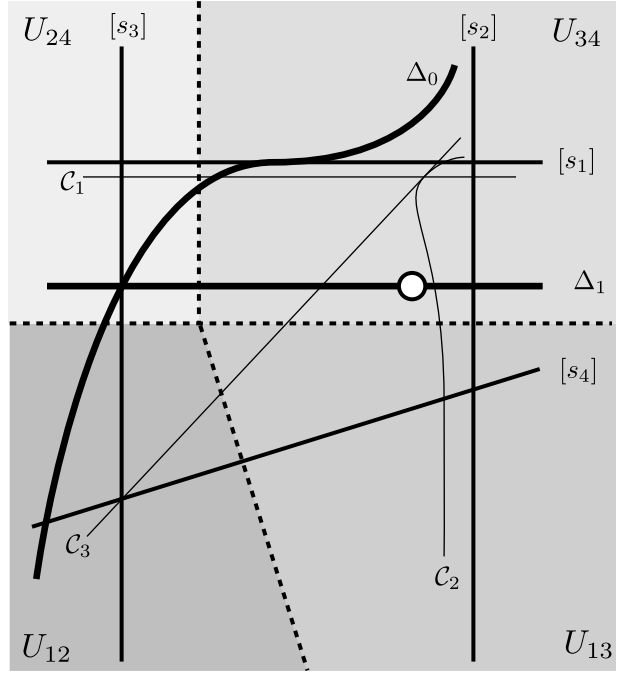}}
   \caption{Schematic depiction of the moduli space of the 2-parameter
   model. Shown are the divisors $[s_i]$ and $\Delta_{0,1}$, and their
   mutual intersections,  the coordinate patches $U_{ij}$ and the
   2-surfaces $\CC_i$ along which our monodromy calculations are done.}
   \label{fig:modspace}
\end{floatingfigure}

To this end, we can choose
\begin{equation}
    \begin{split}
        \CC_1&=\{s_1-\epsilon_1 s_4^3=0\}  \\
        \CC_2&=\{s_2-\epsilon_3\overline{s_4}s_3=0\}  \\
    \CC_3&=\{s_2 s_4-\epsilon_2 s_3=0\}
    \end{split}
\end{equation}

The $\CC_i$ are chosen to meet at our
basepoint near the large radius limit, located at
$(q_1,q_2)=(\epsilon_1,\epsilon_2)$, where
\begin{equation*}
    q_1=\frac{s_1}{s_4^3},\qquad q_2=\frac{s_2s_4}{s_3}
\end{equation*}
are the good local coordinates in the patch $U_{34}$. Note that $\CC_2$ is homotopic to, but is not itself a holomorphic
curve in $\CM$. In the local coordinates in $U_{34}$, it is given by
\begin{equation*}
    C_2=\{q_2=\epsilon_3|s_4(q_1,q_2)|^2\}
\end{equation*}
where $|s_4(q_1,q_2)|^2$ is a complicated
function of $q_1,q_2$, given by solving the D-flatness condition
\eqref{eq:MDflat} and
\begin{equation*}
    \epsilon_3=\frac{\epsilon_2}{|s_4(\epsilon_1,\epsilon_2)|^2}
\end{equation*}

First consider the 2-surface $\CC_1$. It is easy to see
that it intersects $\Delta_0$ three times, $[s_2]$ and $[s_3]$ once each, and
does not intersect $[s_1]$, $[s_4]$ or $\Delta_1$.

The monodromy around the conifold locus is
\begin{equation}
    M_c: v\mapsto v-3(v,\CO)\CO
\end{equation}
As usual, the ``3" is because there are three flat line bundles (6-branes)
which become massless at the conifold locus. The monodromies about the other
two points of intersection with $\Delta_0$ are related to this by conjugation
with $M_{r_1}$,
\begin{equation}
  \begin{split}
    M_{c'}: &\ v\mapsto v-3(v,L_1)L_1\\
          &= M_{r_1} M_c M_{r_1}^{-1}\\
    M_{c''}: &\ v\mapsto v-3(v,L_1^2)L_1^2\\
          &= M_{r_1}^2 M_c M_{r_1}^{-2}\\
  \end{split}
\end{equation}
The monodromy about $[s_3]$, therefore is
\begin{equation}\label{eq:s3monodromy}
   M_{[s_3]}= M_{r_2} M_c M_{c'} M_{c''}
\end{equation}
and satisfies
\begin{equation}\label{eq:s3cubed}
   M_{[s_3]}^3=1
\end{equation}

Similarly, consider $\CC_2$. This intersects
$[s_1]$, $\Delta_1$ and $[s_4]$. The monodromy about $\Delta_1$ is
\begin{equation}
    M_{\Delta_1}:v\mapsto v-(v,x)x
\end{equation}
where $x=\CO_{D_2}-3\CO_{D_1}-3\CO_E$. So the monodromy about $[s_4]$ is
\begin{equation}
   M_{[s_4]}= M_{r_1} M_{\Delta_1}
\end{equation}
and satisfies
\begin{equation}\label{eq:Ms4cubed}
   M_{[s_4]}^3=M_{r_2}
\end{equation}

Finally, we turn to $\CC_3$. Unlike the previous
cases, $\CC_3$, or any 2-surface homotopic to it, necessarily crosses $[s_3]$ at the LG point (the
intersection of $[s_3]$ and $[s_4]$). Thus it makes sense to talk about
the monodromy ``about the LG point''. $\CC_3$ also intersects $[s_1]$, $\Delta_0$
and $\Delta_1$ each once. So we find the monodromy about the $LG$ point is
\begin{equation}
    M_{LG}=M_{r_1}M_c M_{\Delta_1}
\end{equation}
which satisfies
\begin{equation}
    M_{LG}^9=1
\end{equation}

The quantum symmetry at the LG point is enhanced from $\BZ_3$ to
$\BZ_3\times\BZ_9$. The $\BZ_3$ generator is, of course, tensoring with the
flat line bundle $\CL$. The $\BZ_9$ generator is $M_{LG}$. The ($L=0$)
$B$-type fractional branes at the LG point are the orbit under $\BZ_3\times
\BZ_9$ of the D6-brane, $\CO$. They fall into the K-theory classes,
\begin{equation}
    \begin{split}
        V_{k,m}&=V_{k,3}-m\alpha  \\
        V_{k,3}&=M_{LG}^{-k}\ V_{9,3}\\
        V_{9,3}&=\CO
    \end{split}
\end{equation}
or, explicitly,
\begin{equation*}
    \begin{split}
        V_{1,3}&=-2\CO -\CO_{D_1}                  +\CO_E      \\
        V_{2,3}&=  \CO +\CO_{D_1}                         =L_1 \\
        V_{3,3}&=  \CO+3\CO_{D_1} -\CO_{D_2}      +3\CO_E      \\
        V_{4,3}&=-2\CO-7\CO_{D_1}+2\CO_{D_2}+\CO_C-5\CO_E      \\
        V_{5,3}&=  \CO+4\CO_{D_1} -\CO_{D_2}-\CO_C+3\CO_E-\CO_p\\
        V_{6,3}&=-2\CO-3\CO_{D_1} +\CO_{D_2}      -3\CO_E      \\
        V_{7,3}&= 4\CO+8\CO_{D_1}-2\CO_{D_2}-\CO_C+4\CO_E      \\
        V_{8,3}&=-2\CO-5\CO_{D_1} +\CO_{D_2}+\CO_C-3\CO_E+\CO_p\\
        V_{9,3}&=  \CO
    \end{split}
\end{equation*}
The intersection form
\begin{equation}
    (V_{k,m},V_{k',m'})=f_{k-l}
\end{equation}
where $f_n$ takes values
\begin{center}
    \begin{tabular}{|c|ccccccccc|}
        \hline
        $n  $&$ 1 $&$ 2 $&$ 3 $&$ 4 $&$ 5 $&$ 6 $&$ 7 $&$ 8 $&$ 9 $\\
        \hline
        $f_n$&$-1 $&$ 1 $&$-1 $&$ 2 $&$-2 $&$ 1 $&$-1 $&$ 1 $&$ 0 $\\
        \hline
    \end{tabular}
\end{center}

The other limit points of the model are as follows. There is the
aforementioned large radius point (at the intersection of $[s_1]$ and $[s_2]$).
There's a hybrid point, consisting of a LG model (with a cubic superpotential)
fibered over a $\BP^2$, at the intersection of $[s_1]$ and $[s_3]$. Circling
this point about the $[s_3]$, we detected in \eqref{eq:s3cubed} the enhanced
$\BZ_3$ quantum symmetry of the LG fiber; circling this point about $[s_1]$, we
detect the monodromy, $M_{r_1}$, associated to shifting the B-field on the
$\BP^2$ base. Finally, at the intersection of $[s_2]$ and $[s_4]$, the
Calabi-Yau develops a $\BZ_3$ orbifold singularity. The Calabi-Yau isn't
globally a quotient by this $\BZ_3$, so we don't really have an enhanced
$\BZ_3$ quantum
symmetry. Rather, circling $[s_4]$ three times is equivalent to shifting the
B-field \eqref{eq:Ms4cubed}.

\subsection{Branes in the small volume phase}
According to the above discussion, D-branes in the small volume
phase can be investigated by studying D-branes on the orbifold
$\BC^5/\BZ_9\times \BZ_3$. Those can be studied in terms of quiver
gauge theory. A basic set of D-branes (with Dirichlet conditions
in all directions of the orbifold) is  given by the fractional
branes, which are labeled by irreducible representations of
the orbifold theory. Those form the nodes of the quiver. The chiral matter
multiplets, which can be determined in the usual way by projection,
give rise to the links of the quiver. Their number can be computed
from the index theorem and is equal to the intersection number.
This should therefore be compared to the result of a  geometric index
computation.
The continuation of the fractional brane basis to large volume sheaves
has been discussed in the literature
\cite{DiaDou,Mayr,GovJayII,Toma}.
In this paper, our focus
is on the K theory classes and we'd like to compare the fractional
branes to  large volume branes whose K-theory classes are $V_{k,m}$.

Let us make this more concrete for the model at hand.
Since the $\BZ_9\times\BZ_3$ orbifold group is abelian, all irreducible
representations are one-dimensional and can be labeled by two
phases: $\rho=(\exp2\pi ik/9,\exp 2\pi m/3)$.
Working out the representation theory yields the following result:
The number  of chiral multiplets between a brane
$(k,m)$ and a brane $(k',m')$ depends only on the difference
$\Delta k = k-k'$. In particular, it is independent of the $m$ label.
The dependence on $\Delta k$ is summarized in the following table:
\begin{center}
    \begin{tabular}{|c|ccccccccc|}
        \hline
        $\Delta k$ & 1 & 2 & 3 & 4 & 5 & 6 & 7 & 8 & 9  \\
        \hline
         & -1 & 1 & -1 & 2 & -2 & 1 & -1 & 1 & 0  \\
        \hline
    \end{tabular}
\end{center}

Comparison with the table in the previous section shows that
this exactly reproduces the geometrical intersection numbers
of the K-theory classes $V_{k,m}$.

The index computations performed above can be taken to the
IR fixed point of the model, which is described by the Gepner
model. Let us make the connection to boundary CFT results more
explicit.

According to \cite{DiaDou}, the fractional
branes of the quiver discussion
should be directly compared to the set of $L=0$ rational Gepner
boundary states. For the covering theory, the Gepner model
$(k=7)^3 (k=1)^2$, these B-type boundary states have been computed
in \cite{RecSch}. The Gepner model itself is a $\BZ_9$ orbifold of a
tensor product of minimal models (+ other projections, which are
currently not of importance to us),
the $\BZ_9$ being the GSO projection. Accordingly, there is a $\BZ_9$
quantum symmetry, which we denote $g$.
The boundary states are labeled by a single label $M$, $M=0,2,4,\dots 18$,
which can be interpreted as discrete $\BZ_9$ Wilson lines. This
label should be directly compared to the representation label $k$
in the quiver discussion, $M=2k$. The quantum symmetry acts on the boundary
states by $g: M \mapsto M+2$. In geometrical terms this action
maps to the action of
the Gepner monodromy on six-branes $\CO$.

To determine the intersection matrix,
the Witten index ${\rm tr}_R (-1)^F$ has to be evaluated
in the open string Ramond sector.
This is related by a modular transformation to the closed string
amplitude $<M_1| (-1)^{F_L}|M_2>_{RR}$ between boundary states.
To compute the intersection matrix on the covering theory,
the formulas given in \cite{BDLR} can be used.
Due to the symmetry of the model, it can be written in terms
of the shift matrix $g$:
\begin{equation}\label{eq:coverintersection}
I = -3 g^{-1} +3 g^{-2} - 3 g^{-3} + 6 g^{-4} -6 g^{-5} +3g^{-6}
-3 g^{-7} + 3 g^{-8}
\end{equation}

{}From this, the intersection matrix of the
orbifold model can be obtained directly.
The boundary states of
the covering theory are invariant under the $\BZ_3$ orbifold
action. To obtain consistent boundary states of the orbifold
theory, one adds a twisted sector contribution to the boundary
states. This part of the boundary state contains only Ishibashi states
built on fields in the twisted sector.
In  the open string sector they lead to
projection operators,
since the modular transformation of a twisted sector character
leads to an insertion of a group element. The boundary states
are distinguished by $\BZ_3$ representations, which determine
how the projections act in the open string sector.

The index in the orbifold model can be determined without explicit
knowledge of that boundary state. It is sufficient to know that
the orbifold acts freely, which means that there are no RR ground
states in the twisted sector. Therefore, the twisted part of the
boundary state  cannot give rise to new contributions to the
index. All that happens is that there is a projection in the open
string sector, picking out an invariant combination of the
R-ground states counted in \eqref{eq:coverintersection}. To write
the new intersection matrix, we introduce the operation $h$, which
is the quantum symmetry corresponding to the $\BZ_3$. In terms of
the two quantum symmetry operators the intersection matrix reads:
\begin{equation}
I =  \left( g^{-1} + g^{-2} -  g^{-3} + 2 g^{-4} -2 g^{-5} +g^{-6}
+  g^{-7} +  g^{-8} \right) (1+h+h^2).
\end{equation}
This matrix is just a different form of presenting the contents
of the table in the quiver-based discussion.

The form of the intersection matrix shows that transforming
a fractional brane by $h$ cannot change the $\BZ$-valued
RR charge, but only the torsion charge.

We are now ready to apply our general considerations of section \ref{sec:beyond}
to construct a torsion brane as a bound state of BPS states.

There are two ways to do so: One way is to take a fractional brane
and its $h$-transformed anti-brane. The general discussion of
section \ref{sec:beyond} applies to this example, and this system
is tachyon-free. It therefore presents a classically stable state
carrying only torsion charge.

Another way is to take a superposition of fractional branes in
the following way:
\begin{equation}
|T>=h|B> +\sum_{n=1}^8 g^n |B>,
\end{equation}
where $|B>$ is any of the fractional branes.
There are tachyons propagating between the individual branes,
making this configuration unstable. Mapping the brane charges
to large volume shows explicitly that there is a net torsion
charge and the decay product is therefore non-trivial.

\section{The Self-Mirror Example}\label{sec:FHSV}
So far, all of our examples have had $H^4(X)_{tor}=0$. This had several
simplifying consequences. First, we had that the torsion subgroup
$K^0(X)_{tor}=H^2(X)_{tor}$. That is, torsion elements of the K-theory were just
labeled by their (torsion) first Chern classes. Second, the quantum symmetry
group acted trivially on $K^0(X)_{tor}$, because tensoring with a (flat) line
bundle does not change the first Chern class of an object of rank zero.
\begin{floatingtable}{
\begin{tabular}{|c|c|}
\hline
H$_{6}$  & $1$ \\
H$_{5}$  & $\sigma^{{\oplus2}}$ \\
H$_{4}$  & $1\oplus\sigma^{{\oplus2}}\oplus R^{\oplus10}$ \\
H$_{3}$  & $1^{\oplus4}\oplus R^{\oplus20}$ \\
H$_{2}$  & $1\oplus\sigma^{{\oplus2}}\oplus R^{\oplus10}$ \\
H$_{1}$  & $\sigma^{{\oplus2}}$ \\
H$_{0}$  & $1$ \\
\hline
\end{tabular}}
\caption{Action of $\BZ_{2}$ on the homology of $K3\times T^{2}$.}
\label{tab:K3timesT2}
\end{floatingtable}

Further, because $H^3(X)_{tor}$ also vanished (by Poincar\'e
duality), there was no possibility for adding topologically
nontrivial discrete torsion. Even in cases (such as the second
example of \cite{Brunner-Distler}) where the orbifold conformal
field theory admitted discrete torsion, the resulting CFT was
continuously connected to the CFT without discrete torsion. They
lay in the same connected component of the moduli space.

Finally, we had the property that the monodromies in the K\"ahler moduli
space acted trivially on $K^1(X)$ (which is why, heretofore, we have mostly
talked about $K^0(X)$), while the monodromies in the complex structure
moduli space acted trivially on $K^0(X)$.

To see some of the possibilities when $H^4(X)_{tor}\neq0$, we turn to the
Calabi-Yau example of \cite{FHSC}.
Here $X=(K3\times T^2)/\BZ_2$ where the
$\BZ_2$ acts as $(-1)$ on the $T^2$ and as a freely-acting holomorphic
involution of the
$K3$ (under which the holomorphic 2-form is necessarily odd).

Without the $T^2$,
the quotient $K3/\BZ_2$ would be an Enriques surface; \emph{with} the $T^2$, we
obtain a Calabi-Yau which has the structure of a $T^2$ bundle over the Enriques
surface. The holonomy group is $SU(2)\times Z_2$ which, being smaller than
$SU(3)$, means that the fundamental group is not finite. Rather,
$\pi_1(X)=\BZ_2\semidir(\BZ\times\BZ)$, where $\BZ_2: (n,m)\in\BZ\times\BZ \to
(-n,-m)$.

The commutator subgroup $[\pi_{1}(X),\pi_1(X)]=\BZ^{2}$ of elements
of the form $(0,(2m,2n))$. The quotient
\begin{equation}\label{eq:H1Ferrara}
    \homo{1}{X}=\pi_1(X)/[\pi_1(X),\pi_1(X)]=\BZ_2^3
\end{equation}

The involution acts on the homology of $K3\times T^{2}$ as
in Table~\ref{tab:K3timesT2}, where $1$ is the trivial
representation, $\sigma$ is the sign representation and $R$ is the
regular representation (which, over the integers, is irreducible).

\subsection{Computation of the K-theory}
Since the cohomology of $X$ was computed in \cite{Asp95}, we will just
hit the high points of the computation. The $E^2$ term of the CLSS is
\begin{equation*}
  E^2_{p,q}=\homo{p}{\BR P^\infty,\CH_q(K3\times T^2)}
\end{equation*}
where the trivial representation, $1$, leads to the ordinary homology of
$\BR P^\infty$,
\begin{equation*}
    \homo{n}{\BR P^\infty,\BZ}=
      \begin{cases}
      \BZ & n=0\\
      \BZ_2 & n=2k+1,\quad k\geq 0\\
      0 &\text{otherwise}
      \end{cases}
\end{equation*}
The sign representation, $\sigma$, leads to the homology with twisted
coefficients,
\begin{equation*}
    \homo{n}{\BR P^\infty,\widetilde\BZ}=
      \begin{cases}
      \BZ_2 & n=2k\\
      0 &\text{otherwise}
      \end{cases}
\end{equation*}
and the regular representation to
\begin{equation*}
    \homo{n}{\BR P^\infty,\CR}=
      \begin{cases}
      \BZ & n=0\\
      0 &\text{otherwise}
      \end{cases}
\end{equation*}
Putting these together with Table~\ref{tab:K3timesT2}, yields the
$E^2$ term,
\begin{subequations}\label{eq:E2Ferrara}
\begin{center}\qquad\qquad $E^{2}_{p,q}$
  \begin{tabular}{c|cccccc}
    6&$        \BZ          $&$ \BZ_2 $&$   0   $&$ \BZ_2 $&$   0   $&\\
    5&$       \BZ_2^2       $&$   0   $&$\BZ_2^2$&$   0   $&$\BZ_2^2$&\\
    4&$\BZ^{11}\oplus\BZ_2^2$&$ \BZ_2 $&$\BZ_2^2$&$ \BZ_2 $&$\BZ_2^2$&\\
    3&$      \BZ^{24}       $&$\BZ_2^4$&$   0   $&$\BZ_2^4$&$   0   $&$\dots$\\
    2&$\BZ^{11}\oplus\BZ_2^2$&$ \BZ_2 $&$\BZ_2^2$&$ \BZ_2 $&$\BZ_2^2$&\\
    1&$      \BZ_2^2        $&$   0   $&$\BZ_2^2$&$   0   $&$\BZ_2^2$&\\
    0&$        \BZ          $&$ \BZ_2 $&$   0   $&$ \BZ_2 $&$   0   $&\\
    \hline
      & 0 & 1 & 2 & 3 & 4&$\dots$
  \end{tabular}
\hfill\eqref{eq:E2Ferrara}
\end{center}
\end{subequations}

The differential $d_2$ vanishes, but $d_3$ is nontrivial. The
spectral sequence converges at the $E^4$ term, which looks like,
\begin{subequations}\label{eq:E4Ferrara}
\begin{center}\qquad\qquad $E^{4}_{p,q}$
  \begin{tabular}{c|cccccc}
    6&$        \BZ        $&$   0   $&$   0   $&$ 0 $&$ 0 $&\\
    5&$         0         $&$   0   $&$   0   $&$ 0 $&$ 0 $&\\
    4&$\BZ^{11}\oplus\BZ_2$&$ \BZ_2 $&$ \BZ_2 $&$ 0 $&$ 0 $&\\
    3&$      \BZ^{24}     $&$\BZ_2^2$&$   0   $&$ 0 $&$ 0 $&$\dots$\\
    2&$\BZ^{11}\oplus\BZ_2$&$ \BZ_2 $&$ \BZ_2 $&$ 0 $&$ 0 $&\\
    1&$      \BZ_2^2      $&$   0   $&$\BZ_2^2$&$ 0 $&$ 0 $&\\
    0&$        \BZ        $&$ \BZ_2 $&$   0   $&$ 0 $&$ 0 $&\\
    \hline
      & 0 & 1 & 2 & 3 & 4&$\dots$
  \end{tabular}
\hfill\eqref{eq:E4Ferrara}
\end{center}
\end{subequations}

So we find a filtration of $\homo{1}{X}$ which gives
\begin{equation*}
    0\to\BZ_2^2\to\homo{1}{X}\to\BZ_2\to 0
\end{equation*}
We have already computed in \eqref{eq:H1Ferrara} that the extension must
be trivial. We also find directly that
\begin{equation*}
    \homo{2}{X}_{tor}=\BZ_2
\end{equation*}
So we have
\begin{equation*}
    \begin{aligned}
        \coho{2}{X}&=\BZ^{11}\oplus\BZ_2^3\\
        \coho{4}{X}&=\BZ^{11}\oplus\BZ_2
    \end{aligned}
    \qquad
    \begin{aligned}
       \coho{3}{X}&=\BZ^{24}\oplus\BZ_2\\
       \coho{5}{X}&=\BZ_2^3
    \end{aligned}
\end{equation*}

We can choose a basis for $\coho{2}{X}$ as follows. Let the index $I$
run over ten values, $I=+,-,i=1,\dots,8$. We have generators,
$\xi_0,\xi_I$, as well as the torsion generators, $\chi_0,\chi_A$, for
$A=1,2$.
For $\coho{4}{X}$, we choose: $\eta^0,\eta^I$ and the torsion
generator $\phi$. Finally, let $\rho$ be the generator of $\coho{6}{X}$.
The ring structure on $\coho{ev}{X}$ is
\begin{equation}\label{eq:FerraraHevRing}
    \begin{aligned}
        \xi_I\cup\xi_J&=-2C_{IJ} \eta^0\\
        \xi_I\cup\xi_0&=-2C_{IJ}\eta^J
    \end{aligned}
    \qquad
    \begin{aligned}
       \xi_0\cup\chi_0&=\phi\\
       \chi_1\cup\chi_2&=\phi
    \end{aligned}
    \qquad
    \begin{aligned}
       \xi_0\cup\eta^0&=\rho\\
       \xi_I\cup\eta^J&=\delta_I{}^J\rho
    \end{aligned}
\end{equation}
where $C_{IJ}$ is a symmetric matrix, whose nonzero entries are $C_{ij}$,
the Cartan matrix of $E_8$, and $C_{+-}=C_{-+}=-1$.

For $\coho{3}{X}$, we can choose a basis: $\zeta_{AI'},\zeta_{Aa}$, for
$a=1,2$,
and torsion generator $\sigma$. $\coho{5}{X}$ is pure torsion, with generators:
$\kappa_A,\kappa_0$.

The ring structure\footnote{The ring structure is more easily understood
from the Hochschild-Serre Spectral sequence (for the \emph{cohomology} of $X$),
which preserves the multiplicative structure. The $E_4=E_\infty$ term is
\begin{center}\qquad\qquad $E_{\infty}^{p,q}$
  \begin{tabular}{c|ccc}
    6&$        \rho          $&$    0   $&$    0   $\\
    5&$         0            $&$    0   $&$    0   $\\
    4&$  \eta^I,\eta^0       $&$\kappa_0$&$    0   $\\
    3&$\zeta_{AI'},\zeta_{Aa}$&$    0   $&$\kappa_A$\\
    2&$   \xi_I,\xi_0        $&$ \sigma $&$  \phi  $\\
    1&$         0            $&$ \chi_A $&$    0   $\\
    0&$        \BZ           $&$    0   $&$ \chi_0 $\\
    \hline
      & 0 & 1 & 2
  \end{tabular}
\end{center}
where, in each $E_\infty^{p,q}$, we have listed the corresponding generator
of $\coho{\bullet}{X}$ (all of the extensions in the filtration of the
associated-graded being trivial).} is
\begin{equation}\label{eq:HoddRing}
    \begin{aligned}
        \zeta_{AI'}\cup\zeta_{BJ'}=-2\epsilon_{AB}C_{I'J'}\ \rho\\
        \zeta_{Aa}\cup\zeta_{Bb}=\epsilon_{AB}(\sigma_1)_{ab}\ \rho
    \end{aligned}
    \qquad
    \begin{aligned}
        \xi_0\cup\sigma&=\kappa_0\\
       \chi_A\cup\sigma&=\kappa_A
    \end{aligned}
\end{equation}
where $C_{I'J'}$ is the same $10\times10$ matrix as the one which
appeared in \eqref{eq:FerraraHevRing} and $\sigma_1$ is the Pauli
matrix.

Since $\coho{3}{X}_{tor}=\BZ_2$, we have the possibility of turning
on a topologically-nontrivial flat $B$ field, with
$[H]\in\coho{3}{X}_{tor}=\BZ_2$. The moduli space has two disconnected
components\footnote{In fact, there are further possibilities,
involving turning on flat, but topologically-nontrivial RR gauge
fields. The full story, including the dual heterotic description,
would take us too far afield, and will be
discussed elsewhere \cite{Gomis}.}, depending on whether we turn on a
nontrivial $[H]$. If we do so, D-brane charge takes
values in the twisted K-theory, $K^{\bullet}_{[H]}(X)$. The $E_2$ term of
the AHSS is exactly the same as in the untwisted case; only the differentials
are modified. For our purposes, it suffices to know that
\begin{equation}
   d_3 = Sq^3 + [H]
\end{equation}
In our previous paper, we showed rather generally that, for a 6-manifold
$X$ with $H^1(X)=0$, all of the higher differentials in the AHSS vanish. Since
our argument did not invoke the specific form of $d_3$ (merely that its image
is torsion), it works just as well when $[H]$ is a nonzero torsion
element as when it vanishes.

So, in both the twisted and the untwisted cases, we have
\begin{equation}\label{eq:K1SeqFerrrara}
    0\to \coho{5}{X}\to K^1(X)\to \coho{3}{X}\to 0
\end{equation}
for $K^1(X)$ and
\begin{equation}\label{eq:K0torextension}
    0\to \coho{4}{X}_{tor}\to K^0(X)_{tor}\to \coho{2}{X}_{tor}\to 0
\end{equation}
for the torsion in $K^0(X)$. We need to decide whether the extension
\begin{equation*}
    0\to \BZ_2\to K^0(X)_{tor}\to \BZ_2^3\to 0
\end{equation*}
is trivial ($K^0(X)_{tor}=\BZ_2^4$) or nontrivial ($K^0(X)=\BZ_2^2\oplus\BZ_4$).
That is, we want to know if there is an element of order 4 in the torsion
subgroup.

In fact, it is easy to see that no elements of $K^0(X)_{tor}$ are
order 4. We can explicitly construct the generators
\begin{equation}\label{eq:torsionGens}
    \begin{split}
        \alpha_A&=\CL_A-\CO  \\
        \alpha_0&=\CL_0-\CO
    \end{split}
\end{equation}
where $\CL_A$ and
$\CL_0$ are the flat line bundles with first Chern class $\chi_A$
and $\chi_0$, respectively. The remaining generator is
\begin{equation}
    \tilde\alpha=\CL_0\otimes L+\CO-\CL_0-L=a_0\otimes\alpha_0
\end{equation}
where $L$ is the line bundle with $c_1(L)=\xi_0$ and $a_0=L-\CO$.
This generator has $c_1(\tilde\alpha)=0$, $c_2(\tilde\alpha)=\phi$. Since
$\phi$ is
not the square of some class in $\coho{2}{X}_{tor}$, $\tilde\alpha$ cannot  be
written as
\emph{twice} some linear combination of the other generators,  which is what we
would have if
\eqref{eq:K0torextension} were a  nontrivial extension.

\subsection{The moduli space}
The vector multiplet and hypermultiplet moduli space is
\begin{equation*}
 \begin{split}
  \CM_V&=\homoquot{O(10,2)}{O(10)\times (2)}{\Gamma_V}\times\frac{SL(2)}{U(1)}\\
  \CM_H&=\homoquot{O(12,4)}{O(12)\times O(4)}{\Gamma_H}
 \end{split}
\end{equation*}
The modular group $\Gamma_{V}\times \Gamma_H$ is roughly the subgroup of the
modular group of $K3\times T^2$ compactifications which survives the
orbifold projection. We will discuss the more precise definition
in the following; it will depend on whether certain RR fluxes are turned on.

The symplectic form on $K^{1}(X)/K^1(X)_{tor}$ coincides with the standard
intersection form on $\coho{3}{X}/$torsion. That is,
\begin{equation}\label{eq:HoddInt}
   \Omega=(-2C\oplus\sigma_1)\otimes(i\sigma_2)
\end{equation}
where $C_{I'J'}$ is the matrix in\eqref{eq:HoddRing} and the
$\sigma_i$ are the Pauli matrices.

A similar result holds for the intersection form on $K^0(X)$, which can
be best understood as follows. Let  $\pi:X\to\CE$ be the projection from $X$ to
the Enriques surface, $\CE$. Any element $v\in K^0(X)$ can be uniquely
decomposed as
\begin{equation}\label{eq:K0Decomp}
    v= \pi^! u + (L-\CO)\otimes \pi^!\tilde u +  c^A \alpha_A
\end{equation}
where, as above, $L$ is the line bundle with $c_1(L)=\xi_0$,
 $u,\tilde u\in K^0(\CE)$ and the $c^A=0$ or $1$. Representing
$v$  by the quadruple $(u,\tilde u,c^1,c^2)$ and $w$ by the quadruple
$(x,\tilde x,d^1,d^2)$, a simple computation yields
\begin{equation}
    (v,w)= Q(\tilde u,x)-Q(u,\tilde x)
\end{equation}
where $Q(x,y)$ is the \emph{symmetric} quadratic form on $K^0(\CE)$ given by
taking the Dolbeault index on $\CE$,
\begin{equation}
   Q(x,y)=Ind_\CE\overline{\partial}_{x\otimes\overline{y}}
\end{equation}
For later use, it will be helpful to tabulate this quadratic form in some
explicit basis for $K^0(\CE)$ (modulo torsion). Choose $\CO,x_I=L_I-\CO$ and
$\CO_p$ as a basis. Then
\begin{equation}\label{eq:Qbasis}
 \begin{aligned}
   Q(\CO,\CO)&=1\\
   Q(\CO,x_I)&=-C_{II}\\
   Q(\CO,\CO_p)&=1
 \end{aligned}
 \qquad
 \begin{aligned}
   Q(x_I,x_J)&=2C_{IJ}\\
   Q(x_I,\CO_p)&=0\\
   Q(\CO_p,\CO_p)&=0
 \end{aligned}
\end{equation}

As we said, $X$ is a $T^2$-bundle over an Enriques surface, $\CE$. $T^2$ has a T-duality
group $SL(2,\BZ)\times SL(2,\BZ)$. Doing fiber-wise T-duality is a symmetry of
the theory. One of these $SL(2,\BZ)$'s becomes a subgroup of $\Gamma_V$; the
other is a subgroup of $\Gamma_H$. Which is which depends on whether we are
studying Type IIA or Type IIB on $X$.

The modular group $\Gamma_V=O(10,2,\BZ)\times
SL(2,\BZ)$. In the Type IIB description, where the vector multiplet moduli
space is the space of complex structures, the $SL(2,\BZ)\subset\Gamma_V$ is
the ``geometrical" one, acting on $\coho{1}{T^2}$, {\it i.e.}~the one which
acts on the  `$A$' index. In the Type IIA description, the $SL(2,\BZ)\subset
\Gamma_V$ is the one which permutes $\coho{0}{T^2}$ and $\coho{2}{T^2}$.

 Modulo the (torsion) subgroup of $K^0(X)$
generated by the $\alpha_A$ (and the corresponding subgroup of $K^1(X)$
generated by $\beta_A$, the images of the $\kappa_A$ in
\eqref{eq:K1SeqFerrrara}), this
$SL(2,\BZ)$ is generated by
\begin{equation}
 \begin{split}
    T&: v\mapsto v\otimes L\\
    S&: v\mapsto v\otimes (L-\CO) -\pi^!\pi_!v
 \end{split}
\end{equation}
The expression for $S$ is not quite correct as an action on all of
$K^\bullet(X)$, since it annihilates the
$\alpha_A$ and the $\beta_A$. More correctly, $S$ acts on $K^0(X)$ as
\begin{subequations}\label{eq:KahlerSL2Z}
\begin{equation}
  S: (u,\tilde u, c^1,c^2)\mapsto (-\tilde u,u,c^1,c^2)
\end{equation}
and on $K^1(X)$ as
\begin{equation}
  S: \beta_0\leftrightarrow \tilde\beta
\end{equation}
leaving the other generators fixed.
Here $\beta_0$ is the image of $\kappa_0$ in \eqref{eq:K1SeqFerrrara} and
$\tilde\beta$ is the pullback of the torsion class in $K^1(\CE)$. In the same
notation, the action of $T$ is
\begin{equation}
  \begin{split}
    T:\ (u,\tilde u, c^1,c^2)&\mapsto (u,\tilde u + u,c^1,c^2)\\
        \tilde\beta&\mapsto\tilde\beta+\beta_0
  \end{split}
\end{equation}
\end{subequations}
leaving the rest of $K^1(X)$ fixed.

 It is easy to see that $S,T$ satisfy the
desired relations
\begin{equation}
\begin{split}
  S^2&=-1\\
  (ST)^3&=1
\end{split}
\end{equation}
and preserve both the intersection pairing and the torsion pairing.

More subtly, they also commute with the $\BZ_2$ quantum symmetry (when one views
the theory on $X$ as a $\BZ_2$ orbifold of the theory on $K3\times T^2$). The
action of the quantum symmetry is generated by
\begin{equation}\label{eq:FHSVquantSym}
    v\mapsto v\otimes\CL_0
\end{equation}
Clearly, this commutes with the action of $T$. In the above decomposition, it
acts as
\begin{equation}\label{eq:FHSVquantSymII}
  (\BZ_2)_Q: (u,\tilde u, c^1,c^2)\mapsto (u\otimes \CL, \tilde u\otimes
\CL,c^1,c^2)
\end{equation}
where $\CL$ is the flat nontrivial line bundle on the Enriques surface, and it
acts trivially on $K^1(X)$. Thus it also commutes with the action of $S$.

Note that the first $SL(2,\BZ)$ acts trivially on
$K^0(X)/K^0(X)_{tor}$  and the second acts trivially on $K^1(X)/K^1(X)_{tor}$
But \emph{both} act nontrivially on the torsion subgroups. The
torsion elements, $\alpha_A=\CL_A-\CO\in K^0(X)_{tor}$ and $\beta_A\in
K^1(X)_{tor}$, transform as doublets under the first
$SL(2,\BZ)$.

Under the second $SL(2,\BZ)$, the torsion classes $\alpha_0$ and
$\tilde\alpha$ transform as a doublet as do torsion branes,
$\beta_0,\tilde\beta\in K^1(X)_{tor}$.

Note that this is exactly the situation anticipated in
\cite{Brunner-Distler}. It is simply \emph{not true} that $K^0$ is held fixed when
we move about in the complex structure moduli space, and $K^1$ is held
fixed as we move about in the K\"ahler moduli space. Rather, \emph{both}
undergo monodromies. Only after modding out by the torsion do we find
that  $K^0/K^0_{tor}$ is held fixed when
we move about in the complex structure moduli space, and $K^1/K^1_{tor}$ is held
fixed as we move about in the K\"ahler moduli space.

\subsection{Fluxes}\label{sec:fluxes}
In \cite{FHSC}, the authors argue that, to obtain a simple heterotic
dual, one needs to turn on certain RR fluxes. In the Type IIA
description (so that the fluxes are elements of $K^0(X)$), we can turn
on a flux in class $\alpha_0$. That is, we consider turning on a
$\BZ_2$ Wilson line for the RR gauge field.

Under the action of the modular group, $\alpha_0$ is not invariant. Its
orbit under
the $SL(2,\BZ)\subset \Gamma_V$ consists of three elements,
$\alpha_0, \tilde\alpha$ and $\alpha_0+\tilde\alpha$. Before modding out
by $\Gamma_V$, the moduli space consists of four disconnected components,
one with no RR flux turned on, and three more with the above RR fluxes turned
on. The subgroup $\Gamma(2)\subset SL(2,\BZ)$ preserves these RR fluxes
and the quotient group $SL(2,\BZ)/\Gamma(2)$
identifies the different components. The upshot is that the moduli space
with RR flux turned on, rather than having three disconnected components, has
a single connected component
\begin{equation}
    \CM_V^{RR} =
    \homoquot{O(10,2)}{O(10)\times O(2)}{\tilde\Gamma_V}\times\frac{SL(2)}{U(1)}
\end{equation}
where
\begin{equation}\label{eq:newModular}
    \tilde\Gamma_V=O(10,2,\BZ)\times \Gamma(2)
\end{equation}
In particular,
$\CM_V^{RR}$ is a finite cover of the previously-discussed vector multiplet
moduli space with no RR flux turned on.

In \cite{deBoer:Triples}, it was argued that turning on a discrete RR flux
$f\in K^\bullet(X)_{tor}$  restricts the allowed D-brane charges in the
theory.  It is a little hard to directly compare their results to ours, as
they are interested in the equivariant K-theory of orbifolds which \emph{cannot}
be resolved to smooth manifolds.
The condition they proposed was that only those charges, $v$,
which satisfy
\begin{equation}\label{eq:deBoer}
    f\otimes v=0
\end{equation}
are allowed. The set of such $v$'s forms a subgroup of $K^\bullet(X)$.
As we will see in the next section, this proposal does not seem to give
the right answer in the case we are interested in. Instead, we propose that
the correct group of D-brane charges is
\begin{subequations}\label{eq:us}
\begin{equation}
    K^\bullet(X)/\Gamma_f
\end{equation}
where $\Gamma_f$ is the (torsion) subgroup of $K^\bullet(X)$ given by
\begin{equation}
   \Gamma_f =\{ v \in K^\bullet(X)\
   \text{such that}\ v=f\otimes w\ \text{for some}\ w\}
\end{equation}
\end{subequations}

In the present instance, $f=\alpha_0$, and the restriction \eqref{eq:deBoer}
is that the
coefficient of $a_0=L-\CO$ in the charge must be \emph{even}. There is
no restriction on the charges in $K^1(X)$.  In terms of the
decomposition $v=(u,\tilde u,c^1,c^2)$ of \eqref{eq:K0Decomp}, it means that
the element $\tilde u\in K^0(\CE)$ has \emph{even} rank. And, indeed,
the $\Gamma(2)$ subgroup of $SL(2,\BZ)$ preserves\footnote{Actually,
\eqref{eq:deBoer} is preserved by the larger
subgroup, $\Gamma_0(2)\subset SL(2,\BZ)$. If \eqref{eq:deBoer} is the
right condition, it would be more natural for the modular group to be
$\tilde\Gamma_V=O(10,2,\BZ)\times \Gamma_0(2)$ rather
than \eqref{eq:newModular}. It is only because we
require that \eqref{eq:us} to be preserved that we insisted on
\eqref{eq:newModular} above.} this condition on
$\tilde u$.

The restriction \eqref{eq:us} in our case is less drastic. The image of
$f\otimes$ is
\begin{equation}
    \Gamma_f = \{0,\tilde\alpha\}
\end{equation}
and our proposal for the group of D-brane charges is the quotient of the
K-theory by this $\BZ_2$ subgroup.

In terms of the decomposition  $v=(u,\tilde u,c^1,c^2)$, this quotient is simply
expressed by saying that $\tilde u\in K^0(\CE)/K^0(\CE)_{tor}$. The
quantum symmetry \eqref{eq:FHSVquantSymII} should, then, be thought of as acting
by
\begin{equation}\label{eq:FHSVquantSymIII}
  (\BZ_2)_Q: (u,\tilde u, c^1,c^2)\mapsto (u\otimes \CL, \tilde u,c^1,c^2)
\end{equation}

The quantum symmetry in this Type IIA description seems to
be related in a simple, but nontrivial way to the quantum symmetry of
the heterotic dual theory,
\begin{equation*}
    (\BZ_2)_Q^{HET} = (\BZ_2)_Q \cdot (-1)^{\text{rank}(\tilde u)}
\end{equation*}

\subsection{Singularities}\label{sec:Singularities}
\subsubsection{$\CN=4$ degenerations}
Recall that $X$ is a $T^2$ fiber bundle over the Enriques surface, $\CE$. Say,
in Type IIA, we tune the K\"ahler moduli so that a genus-zero curve in the base,
$C\subset
\CE$, shrinks to zero size. The local geometry of $X$ at such a singularity
is just $T^2\times (\BC^2/\BZ_2)$. This local geometry preserves $d=4$,
$\CN=4$ supersymmetry. D2-branes wrapping the curve $C$ become massless in this
limit and give rise to an $\CN=4$ $SU(2)$ gauge theory \cite{FHSC}. That is,
quantizing these D2-branes yields (in $\CN=2$ language) massless vector
multiplets and a massless hypermultiplet in the adjoint.

The monodromy about this locus is easy to compute. From
\eqref{eq:Qbasis}, we have $Q(\CO_C,\CO_C)=4$ and $Q(u,\CO_C)\in2\BZ$ for
\emph{any} $u\in K^0(\CE)$. The monodromy is
\begin{equation}\label{eq:FHSVADE}
   M_C: (u,\tilde u,c^1,c^2)\mapsto
   (u-\tfrac{1}{2}Q(u,\CO_C)\CO_C,
     \tilde u-\tfrac{1}{2}Q(\tilde u,\CO_C)\CO_C,c^1,c^2)
\end{equation}
By the above remark, $\tfrac{1}{2}Q(x,\CO_C)$ is always an integer, so the
above formula makes sense. As expected, since the singular locus in question is
a $\BZ_2$ orbifold locus in the moduli space, $M_C$ satisfies
\begin{equation}
  M_C^2=1
\end{equation}

If you wish, you can cast \eqref{eq:FHSVADE} in the form of
\eqref{eq:KMP}, for $n=0$, by taking
\begin{equation*}
  \begin{split}
    y& =(0,\CO_C,0,0)\\
    x\ ``&{=}"\ \tfrac{1}{2}(\CO_C,0,0,0)
  \end{split}
\end{equation*}
If we were considering $K3\times T^2$, instead of its quotient, $X$, we would
have $Q(\CO_C,\CO_C)=2$ and we could write down an honest formula of the form
\eqref{eq:KMP}, with no pesky factors of $\tfrac{1}{2}$.

The attentive reader will note that the monodromy \eqref{eq:FHSVADE} does not
commute with the $SL(2,\BZ)$ action \eqref{eq:KahlerSL2Z} (or even with
its $\Gamma(2)$ subgroup). This should be
obvious from the physical description that we have given of the singularity.
The $SL(2,\BZ)$ of \eqref{eq:KahlerSL2Z} mixes D2-branes wrapped on $C$ with D4-branes wrapped on $T^2\times
C$. \eqref{eq:FHSVADE} describes the monodromy along a path circling the complex
codimension-one locus where the D2-brane wrapped on $C$ becomes massless.
Conjugating $M_C$ with $S$ in (\ref{eq:FHSVADE}a), we obtain the monodromy
along a \emph{different} path through the moduli space: the
one such that, at the singularity, D4-branes wrapped on $T^2\times C$ become
massless.

At higher codimension in the moduli space, we obtain $\CN=4$ ADE singularities,
of a form that should now be quite familiar.

\subsubsection{$\CN=2$ degenerations}
As studied in \cite{FHSC,Asp95,Harvey-Moore:Threshold}, there
is another class of singularity in the moduli space, associated to the
entire Enriques surface $\CE$ collapsing to zero size. Because the
singularity is no longer localized on $\CE$, the physics of the
singularity is \emph{sensitive} to the fact that the $T^2$ bundle over $\CE$ is
twisted. So, instead of obtaining an $\CN=4$ supersymmetric spectrum
of massless states (a vector multiplet and an adjoint hypermultiplet),
the twisting breaks $\CN=4$ to $\CN=2$ and the massless states are an
$SU(2)$ vector multiplet with $N_f=4$ hypermultiplets in the
fundamental representation \cite{FHSC}.

How does this come about? To understand it, we need to make a little
digression about divisors on $X$, in particular those which, under
the projection $\pi: X \to \CE$, cover the Enriques.

A section of this
fiber bundle would give an embedding of Enriques in $X$, whose image would be
a divisor in $X$ which projects down to a single copy of $\CE$.
Since the fiber bundle is nontrivial, there ``generically'' won't be a
section. However, the transition functions of the fiber bundle act
as $-1$ on the $T^2$ fibers. This has 4 fixed points on $T^2$ and, if
we choose the local section to land at one of these fixed points, the
result pieces together to a global section $i: \CE \hookrightarrow X$.
This gives us four divisors, $D_{i,j}$, $i,j=0,1$, in $X$ labeled by
the fixed points of the $\BZ_2$ action on $T^2$. If we wrap a
D4-brane on one of these divisors, we have a BPS brane with charge
\begin{equation}\label{eq:pushforwardO}
    \begin{split}
        \CO_{D_{0,0}} &= L-\CO  \\
        \CO_{D_{1,0}} &= L-\CO +\alpha_1  \\
        \CO_{D_{0,1}} &= L-\CO +\alpha_2  \\
        \CO_{D_{1,1}} &= L-\CO +\alpha_1 +\alpha_2
    \end{split}
\end{equation}
where the $\alpha_A$ of \eqref{eq:torsionGens} are the torsion elements
of $K^0(X)$ which transform under the ``geometrical'' $SL(2,\BZ)$ action
on the torus (the $SL(2,\BZ)$ which is part of $\Gamma_H$).

These branes had a trivial line bundle on their world-volume. But the
Enriques surface also has a flat, but nontrivial, line bundle, $\CL$,
and we would just as well have gotten a BPS brane by wrapping a
D-brane with $\CL$ on its world-volume\footnote{There is a subtlety
here. The normal bundle of one of these divisors in $X$ is the flat,
but nontrivial line bundle $\CL$. So, in order to wrap a D4-brane on
$D$, we need to choose a $Spin^\BC$ structure on $D$. This was
frequently the case when we wrapped 4-branes on divisors in our other
examples. But there the $Spin^\BC$ structure was unique, and so we
did not bother remarking on it. Here, however, because
$\coho{2}{D}_{tor}=\BZ_2$, we have a choice of two different $Spin^\BC$
structures. With one choice of $Spin^\BC$ structure, pushing-forward
$\CO$ gives \eqref{eq:pushforwardO} while pushing forward $\CL$ gives
\eqref{eq:pushforwardL}. With the other choice, pushing forward $\CO$
gives\eqref{eq:pushforwardL}
and $\CL$ gives \eqref{eq:pushforwardO}.}. These give four more BPS
brane, with charges
\begin{equation}\label{eq:pushforwardL}
    \begin{split}
        \CL_{D_{0,0}} &= L-\CO +\tilde\alpha\\
        \CL_{D_{1,0}} &= L-\CO +\alpha_1  +\tilde\alpha\\
        \CL_{D_{0,1}} &= L-\CO +\alpha_2 +\tilde\alpha\\
        \CL_{D_{1,1}} &= L-\CO +\alpha_1 +\alpha_2 +\tilde\alpha
    \end{split}
\end{equation}
The quantum symmetry \eqref{eq:FHSVquantSym} acts to exchange the
branes \eqref{eq:pushforwardO}  with the corresponding branes
\eqref{eq:pushforwardL}.

Another possibility for finding a divisor in $X$ which covers $\CE$ is
to choose a pair of points $(z,-z)$ in the fiber, which  are
exchanged by the $\BZ_2$ transition functions.  This pieces together
to a divisor, $D_W$ in $X$ which double-covers $\CE$. The normal
bundle to $D_W$ in $X$ is trivial (we can vary $z$), so $D_W$ has
trivial canonical bundle. $D_W$ double-covers $\CE$, so it is a K3
surface. A D4-brane wrapped on $D_W$ has charge
\begin{equation}\label{eq:DW}
    \CO_{D_W} = 2(L-\CO)
\end{equation}
This is \emph{twice} the charge of one of the branes in
\eqref{eq:pushforwardO},\eqref{eq:pushforwardL}. (Multiplying by 2 wipes
out the torsion charge.)

This is \emph{almost} the spectrum of wrapped D-branes that we want.
The branes wrapped on $D_W$ give rise to massive vector multiplets
which, when the Enriques shrinks to zero size, produce an enhanced
$SU(2)$ gauge symmetry. The branes wrapped on the $D_{i,j}$ yield
hypermultiplets in the fundamental of $SU(2)$.

Unfortunately, between \eqref{eq:pushforwardO} and
\eqref{eq:pushforwardL}, we seem to have produced two times too
many of them. But, of course, we have yet to implement the fact
that we have turned on a discrete RR flux.

Turning on the RR flux $f=\alpha_0$ changes the spectrum of allowed D-branes.
In \S\ref{sec:fluxes}, we discussed two proposals,
\eqref{eq:deBoer},\eqref{eq:us}, for what this restriction
might be. Let us see what each of them imply in the present context.

Imposing the condition \eqref{eq:deBoer} has no
effect on the branes wrapped on $D_W$. But it does project out all of
the singly-wrapped branes on the $D_{i,j}$.
If you took a trivial rank-2 bundle on one of the $D_{i,j}$, this
could simply decay into a rank-1 trivial bundle on the double-cover and
move off the fixed point to become a rank-1 trivial bundle on $D_W$.
In other words, that does not correspond to a brane ``stuck'' to this
divisor.
Instead, we can take the flat, but nontrivial rank-2 bundle
$\CO\oplus\CL$ on $D_{i,j}$. This cannot decay to a rank-1 bundle on
$D_W$; it is genuinely stuck on the Enriques.
Unfortunately, it also has twice the charge of a field in the
fundamental ({\it i.e.}~it has the same charge as the $W$ bosons
which come from wrapping a 4-brane on $D_W$).
That is clearly not what the physics requires.

Instead, let us see what \eqref{eq:us} implies. Here we find that we
must mod out by $\Gamma_f=\{0,\tilde\alpha\}$. That is, we should
\emph{identify} the branes \eqref{eq:pushforwardO} with the corresponding
branes \eqref{eq:pushforwardL}. This also gives us four distinct branes
wrapping the different $D_{i,j}$, but this time these branes have the
\emph{right} charges to be in the fundamental representation of $SU(2)$.
Note also that the modular group (the subgroup of $SL(2,\BZ)$ which
commutes with the flavour symmetry) for $N_f=4$ was found by Seiberg and Witten
\cite{Seiberg-Witten:N=2II}
to be $\Gamma(2)$. This, too, is in accord with \eqref{eq:us}, rather
than \eqref{eq:deBoer}, which is invariant under the larger group, $\Gamma_0(2)$.

While we don't have a rigorous proof, we believe that the monodromy about this
locus takes the form
\begin{equation}
   M: (u,\tilde u,c^1,c^2)\mapsto
        (u-2Q(u,\CO)\CO,\tilde u-2Q(\tilde u,\CO)\CO,c^1,c^2)
\end{equation}
This preserves the relevant pairings and gives
the right monodromy in the field theory
limit \cite{Seiberg-Witten:N=2II}.

\section*{Acknowledgements}

We would like to thank P.~Aspinwall,
J.~de Boer, P.~Berglund, R.~Donagi, D.~Freed,
J.~Gomis, S.~Kachru, A.~Klemm and D.~Morrison for discussions.
J.~D.~would like to thank the
organizers of the M-Theory Workshop and the ITP for providing a
stimulating environment while this work was being carried out.
I.~B.~thanks the organizers of the M-Theory Workshop and of the
Duality Workshop at the ITP.

This research was supported in part by the National Science
Foundation under Grant No. PHY99-07949.



\bibliography{torsionII}
\bibliographystyle{utphys}

\end{document}